# A community-led calibration of the Zr isotope Reference Materials: NIST candidate RM 8299 and SRM 3169

François L.H. Tissot[1], Mauricio Ibañez-Mejia[2], Savelas A. Rabb[3], Rebecca A. Kraft[3], Robert D. Vocke[3], Manuela A. Fehr[4], Maria Schönbächler[4], Haolan Tang[5,6] and Edward D. Young[5]

[1]The Isotoparium, Division of Geological and Planetary Sciences, California Institute of Technology, Pasadena, CA 91125, USA (tissot@caltech.edu).

[2]Department of Geosciences, The University of Arizona, Tucson, AZ 85721, USA (ibanezm@arizona.edu).

[3]National Institute of Standards and Technology, Gaithersburg, MD 20899 USA.

[4]Institute of Geochemistry and Petrology, ETH Zürich, 8092 Zürich, Switzerland.

[5]Department of Earth, Planetary, and Space Sciences, UCLA, Los Angeles, 90095, USA.

[6]Department of Earth and Space Sciences, University of Science and Technology of China, Hefei, 230026, China.

**Abstract (~292 words):**

As the field of zirconium (Zr) stable isotopes is rapidly expanding from the study of mass-independent to that of mass-dependent isotope effects, a variety of Zr standards have appeared in the literature. While several of these standards have been proposed as the ideal isotope reference material (iRM) against which all data should be reported, none of them have been shown to meet the compositional and/or conflict-of-interest-free distribution requirements put forth by the community. To remedy this situation, we report on a community-led effort to develop and calibrate a scale defining iRM for Zr isotopes: NIST RM 8299. Developed in partnership with the National Institute of Standards and Technology (NIST) from the widely used SRM 3169 Zirconium Standard Solution (certified for mass fraction), the candidate RM 8299 was calibrated through an inter-laboratory study involving three laboratories. Our data show that candidate RM 8299 meets all requirements of an ideal iRM. It is an isotopically homogeneous, high-purity reference material, that is free of isotope anomalies, and whose composition is identical to that of a major geological reservoir (Ocean Island Basalts). Furthermore, RM 8299 will be curated and distributed by NIST, a neutral, conflict-of-interest free organization, and was produced in sufficient quantities to last multiple decades. We recommend that all Zr isotope data be reported against RM 8299. Our results also show that SRM 3169 lots #130920 and #071226 have indistinguishable composition to candidate RM 8299. Therefore, using RM 8299 as the scale defining iRM will enable direct comparison of all future data with the vast majority of the existing literature data, both for mass-independent and mass-dependent isotope effects. To facilitate conversion of $\delta^{94/90}$Zr values reported against other Zr standards, we provide high-precision conversion factors to the RM 8299 scale obtained using the double-spike method.

**Significance to JAAS (<100 words):**

This study addresses a critical need in the field of zirconium stable isotopes by developing an isotope reference material (RM) that adheres to community-defined best practices: NIST RM 8299. The calibration of this new iRM is highly relevant to JAAS, as it involves comparison of MC-ICP-MS (Multi-Collector Inductively-Coupled-Plasma Mass-Spectrometer) data acquired independently in three different laboratories. Beyond demonstrating the isotopic homogeneity of independently bottled aliquots of RM 8299, the results provide insights into the nature of mass-fractionation in MC-ICP-MS instruments, and highlight the challenges associated with accurate determination of absolute ratios using MC-ICP-MS.

# Main Text

## 1. Introduction

Over the past two decades, the field of stable isotope geo- and cosmochemistry has expanded from the so-called "traditional" systems (H, C, N, O, S; *e.g.*, Ref.[1]) to include most of the periodic table (*e.g.*, Refs.[2–4]). For each new isotope system being developed, establishing a scale defining isotopic reference material (iRM) is a critical task that is, unfortunately, not always given sufficient consideration. This has led to problematic situations where data are reported relative to iRMs that were, for instance, inadequate (*e.g.*, isotopically heterogeneous), not readily available, rapidly exhausted (*e.g.*, JMC Lyon Zn), and/or not broadly agreed upon (*e.g.*, various in-house Mo standards in the literature). As the field of (mass-dependent) zirconium (Zr) stable isotopes is starting to develop (*e.g.*, Refs.[5–12]), it is in the community's best interest to establish and rigorously calibrate an iRM early on, before multiple 'in-house' standards proliferate.

In the published literature, Zr isotope data have already been reported against at least 7 different standards. Study of mass-independent isotope effects have used a single-element Zr solution from Johnson Matthey[13], a standard solution prepared from AMES Zr metal[14], at least two lots of NIST SRM 3169 (Refs.[13,15–23]), as well as 4 different lots of single-element Zr solution from Alfa-Aesar[19,24–27]. An early study showed that different Zr single-element standard solutions had identical isotopic composition at the ~20-120 part-per-million (ppm) level[28]. When proceeding to higher precision (< 10 ppm), however, resolvable isotopic differences were identified between NIST SRM 3169 and Alfa Aesar Zr standards[23,26,27], emphasizing the need for a common iRM to render the data of different publications comparable.

Similarly, in the nascent field of Zr mass-dependent stable isotopes, multiple standards have already been used, such as NIST SRM 3169 (Ref.[26,29–35]), the candidate RM 8299 presented herein[7–9,11,12,36–38], a single-element Zr solution from PlasmaCal (SCP Science) also known as the IPGP-Zr standard[5,6,10,11,30,33,35,38–46], and the NRC-ZIRC 1 (Refs.[44,47,48]). Faced with such a proliferation of standards, it is critical to assess whether any of these materials are a suitable primary iRM against which all isotope data, both mass-dependent and independent isotope effects, should be reported. Doing so requires a careful, community-led, interlaboratory calibration effort.

Here, we introduce the candidate RM 8299 reference material. This Zr iRM was conceived and developed in collaboration with the National Institute of Standards and Technology (NIST) in order to meet all the desirable key criteria of an iRM including isotopic homogeneity, wide availability, conflict-of-interest–free distribution, and sufficient stock to last decades (*Section 2*). To ensure traceability to the vast majority of mass-independent isotope studies, candidate RM 8299 was prepared from the widely used SRM 3169 Zirconium Standard Solution. Below, we describe how the candidate RM 8299 was prepared by NIST, and calibrated through a careful and thorough inter-laboratory study involving three participating laboratories at Caltech, ETH Zürich and UCLA (*Section 3 & 4*). Our results confirm that all candidate RM 8299 aliquots are isotopically homogeneous (at the ± 2.5 part-per-million (ppm) level for $^{91}Zr/^{90}Zr$, and $^{92}Zr/^{90}Zr$ ratios, and ± 7.5 ppm for the $^{96}Zr/^{90}Zr$ ratio) and have a composition identical to SRM 3169 (*Section 5*). The data also reveals the existence of clear non-exponential mass-dependent fractionation occurring in each laboratory, from measurement session to measurement session, as well as mass-independent effects between laboratories (*i.e.*, instrument specific) whose origin remains unsolved. These results further emphasize the need for isotopic measurements to be made using a unique iRM, such that anomalous effects can be readily accounted for by sample-standard bracketing to the same standard, and enable direct inter-lab data comparison at the < 5-10 ppm level for both mass-independent and mass-dependent effects.

## 2. Key features for an iRM

Reference materials, both for mass fraction and isotopic analyses, are critical to ensure data accuracy and comparability between methods and laboratories across the world. But selecting and establishing an iRM is a non-trivial task that is often overlooked and/or not given sufficient consideration. In non-traditional stable isotope geochemistry in particular, groups pioneering a new isotope system have little choice but to use a readily available mass fraction standard solution as an in-house isotope standard. By inertia, this in-house standard often becomes the isotope standard for the community, leading to potential issues in the distribution (*e.g.*, conflict-of-interest between competing groups), preservation, and/or long-term availability of this primary reference material. The latter point is the major recurring issue; stocks of iRMs can rapidly become exhausted, forcing calibration of novel iRMs when little of the original material remains available, which leads to potentially problematic inter-calibration issues and/or a proliferation of materials against which isotope data are reported.

To avoid such problems, the community has developed a series of detailed guidelines and best-practices for the selection of an iRM in isotope geochemistry[49–51], which are briefly summarized in Table 1. The preparation and calibration of the NIST candidate RM 8299 followed these guidelines. To ensure the homogeneity (#1) and purity (#2) of the Zr iRM, the candidate RM 8299 was produced by dilution of the high-purity SRM 3169 Zirconium Standard Solution, lot # 130920. The resulting RM 8299 will be curated and distributed by NIST, a reliable and neutral entity, allowing conflict-of-interest free distribution and wide availability (#6). The large volume of the RM lot (>20 L) was sufficient to produce >1000 units (2 bottles each; 10 mL of solution per bottle). At the current rate of sale of the SRM 3169, the current RM 8299 lot would last more than a century (#7). Establishing the composition (#3) and confirming the lack of isotopic anomalies (#4) in candidate RM 8299 was accomplished via high-precision isotope analyses. Finally, and to ensure that the choice of the iRM was a community-led effort (#5), the team that initially conceived the iRM (Caltech, U. of Arizona, and NIST), invited all groups with expertise in Zr isotopic analysis to participate in an inter-laboratory measurement campaign for candidate RM 8299. As we show below, this approach was not only successful in calibrating the candidate RM 8299, but also led to the identification of systematic discrepancies with measurements of absolute ratios on MC-ICPMS instruments, thereby reemphasizing the need for all laboratories to use the same, well-characterized iRM to ensure inter-lab data comparison.

## 3. Inter-laboratory calibration

The protocol for inter-laboratory calibration (Fig. 1) was designed by the NIST (Statistical Engineering Division) to ensure that the source of any difference in the isotopic compositions measured by different labs, or in different bottles, could be identified without the need for any iterative process (*e.g.*, investigation of more solution aliquots, or additional rounds of analyses). In total, each participating laboratory analyzed 5 separate aliquots of candidate RM 8299. The first 3 aliquots were sent directly from NIST (in three separate bottles) to each laboratory. Upon reception, each laboratory sampled the bottles and sent 2 of them to one of the other laboratories. Once all 5 aliquots had been received, each participating group independently decided on the best way to measure the aliquots: *e.g.*, instrument, sample inlet system, interference monitoring (*Section 4.2*). Measurements of all 5 candidate RM 8299 aliquots were to be performed twice (or more) in each laboratory, during at least two separate analytical sessions over non-successive days. This design allows for identification of any isotopic offsets between aliquots, any occurrence (and source) of contamination in a given bottle, as well as any offset stemming from systematic differences between laboratories (*i.e.*, operator- and/or instrument-specific biases, interference correction, data treatment).

## 4. Methods

### *4.1. iRM preparation and aliquoting*

The Zr NIST candidate RM 8299 was produced by dilution of SRM 3169 Zirconium Standard Solution (lot #130920). This lot of SRM 3169 was initially produced by digestion of a large piece (100 g) of high-purity Zr metal, which was then diluted in a mixture of 10 % vol $HNO_3$ + 2 % vol HF (or ~1.5 mol $L^{-1}$ $HNO_3$ + 0.6 mol $L^{-1}$ HF) in NIST laboratories, to yield a Zr mass fraction of 10.000 ± 0.020 mg/g. To produce the NIST candidate RM 8299, an aliquot of SRM 3169 was taken and diluted by a factor ~100 into a matrix of 2.1 % $HNO_3$ + 0.48 % HF (or ~0.32 mol $L^{-1}$ $HNO_3$ + 0.13 mol $L^{-1}$ HF) to yield a Zr mass fraction of ~97.5 µg/g. The volume of RM produced (>20 L) enabled the production of more than 1000 units; each unit consisting of 2 bottles containing 10 mL of solution each.

For the inter-laboratory comparison, twelve bottles of candidate RM 8299 were selected by stratified random sampling. Each bottle was placed in an individual bag and sealed. Three such bottles were shipped to each participating laboratory.

### *4.2. Isotopic analyses*

#### *4.2.1. Inter-laboratory calibration of NIST candidate RM 8299 and SRM 3169*

Each participating laboratory was tasked with defining their own ideal measurements setup. While all groups used multi-collector inductively-coupled-plasma mass-spectrometer (MC-ICP-MS) instruments (Thermo Fisher Scientific Neptune and Neptune Plus), the sample introduction systems, cup configurations and analytical methods varied between groups. The details of these respective approaches are presented in Table 2.

In Laboratory 1, analyses were run in wet plasma conditions, using a glass spray chamber and 1 µg/g Zr solutions prepared by diluting 60 µg of the 100 µg/g Zr candidate RM 8299 aliquots with 5.94 g of run acid (0.58 mol $L^{-1}$ $HNO_3$ + 0.01 mol $L^{-1}$ HF). The instrument was tuned for optimal sensitivity and stability. Analyses were performed over two independent analytical sessions, one in April and the other in June 2022. For each session, a new batch of run acid solution was prepared, as well as new dilutions of the candidate RM 8299 aliquots. In both sessions, the solution from one of the candidate RM 8299 aliquots (bottle 12) was used as the bracketing standard, and each unknown was bracketed as follows: STD - unknown - STD. Aliquots of the SRM 3169 (lot #130920) were measured alongside the candidate RM 8299 solutions, to assess their level of isotopic homogeneity. Before any solution analysis, the same run acid solution (*i.e.*, acid blank) was measured to determine On Peak Zero (OPZ) intensities. All aliquots were measured 8-10 times per measurement session, and aliquot #12 (used as bracketing standard) was measured 37 times in session 1 and 157 times in session 2 (other pure Zr metals, whose composition will be reported in a subsequent study, were also measured during this second session, and bracketed by aliquot #12).

In laboratory 2, analyses were run in dry plasma conditions, using an Aridus II desolvating nebulizer, and 350 ng/g Zr solutions prepared by diluting 56 μL of the 100 μg/g candidate RM 8299 aliquots with 0.1097 mL of conc. HNO$_3$, 2.97 mL H$_2$O and 12.865 mL of run acid (0.5 mol L$^{-1}$ HNO$_3$ + 0.005 mol L$^{-1}$ HF). The instrument and desolvating nebulizer (Aridus II) were tuned to minimize interferences on all Zr masses (including the ArArO$^+$ interference on $^{96}$Zr). Since the Aridus II tuning has a large influence on the absolute Zr isotope ratio measured, the tuning was further adjusted until the ratios obtained on the NIST SRM 3169 were close to the long-term values measured in this laboratory. Analyses were performed over two independent analytical sessions, one in May and the other in June 2020. In both sessions, a SRM 3169 solution (lot #071226) was used as the bracketing standard. In session 1, each unknown was bracketed by a standard measurement (*i.e.*, STD - unknown - STD), while in session 2, a standard was run every 6 analyses as STD - Aliquot 1 - Aliquot 2 - Aliquot 3 - Aliquot 4 - Aliquot 5 - Aliquot 6 - STD. Before analyses of any solution, the same run acid solution was measured to determine OPZ intensities. All Zr candidate RM 8299 aliquots were measured 6 times per session.

In laboratory 3, analyses were run in wet plasma conditions, using a glass spray chamber and 2 μg/g Zr solutions prepared by diluting 40 μg of the 100 μg/g Zr candidate RM 8299 aliquots with 20 g of run acid (0.3 mol L$^{-1}$ HNO$_3$ + 0.005 mol L$^{-1}$ HF). The instrument was tuned for optimal sensitivity and stability. Analyses were performed over three independent analytical sessions, one in February 2020, and two in July 2020. In all sessions, a SRM 3169 solution (lot #130920) was used as the bracketing standard, and each unknown was bracketed: *i.e.*, STD - unknown - STD. At the beginning of each sequence (~ twice a day) a run acid solution was measured to determine OPZ intensities. All Zr candidate RM 8299 aliquots were measured 24 to 42 times over the three sessions.

### *4.2.2. Mass-dependent ($\delta^{94/90}$Zr) analyses of reference materials*

In the absence of a community agreed-upon Zr iRM, the rapid expansion of investigations of Zr mass-dependent effects has led to the proliferation of proposed RMs. These proposed standards cover a wide range of Zr isotope compositions (expressed in delta notation as $\delta^{94/90}$Zr values). To facilitate conversion of the literature data to the $\delta^{94/90}$Zr$_{RM\ 8299}$ scale, we therefore measured at high-precision the composition of these standards against candidate RM 8299. We also measured the composition of several widely used rock and zircon geostandards. Measurements were made on a Thermo Scientific Neptune Plus MC-ICP-MS at the Isotoparium (Caltech), using a $^{91}$Zr-$^{96}$Zr double spike and following the methods from Ref.[8]. Each proposed reference material was measured between 18 and 597 times over multiple analytical sessions spread over several years.

### *4.3. Data reduction*

To avoid potential systematic biases from differences in data treatment, the data from all laboratories were reduced offline (see *Supplementary Materials*) using a consistent procedure, which is briefly summarized below. First, the raw intensities of all measurements and their associated OPZ values were tabulated. For each measurement, the OPZ values were subtracted from the raw measurement intensities. A correction for isobaric interferences was then implemented. In principle, both molybdenum ($^{92}$Mo, $^{94}$Mo, and $^{96}$Mo) and ruthenium ($^{96}$Ru) isotopes can impact Zr isotope analyses, and interferences from both elements should be corrected for. In practice, the high-purity of the solutions resulted in extremely small interference intensities (at most 30 μV), and the Ru correction was found to be negligible (< 1.5 ppm offset on the internally normalized $^{96}$Zr/$^{90}$Zr ratios). In contrast, the Mo correction was found to have a resolvable impact on the data (see *Section 6.2.1*). In all cases, the beam intensities of the interfering species were subjected to a first-order instrumental mass-bias correction, using the exponential law[52] and internal normalization to a $^{94}$Zr/$^{90}$Zr (calculated from OPZ-corrected intensities) of 0.3381 (Ref.[53]), for consistency with the more extensive literature on Zr mass-independent isotope variations. The correction of isobaric interferences therefore used the following general equation:

$$^{9x}Zr_{Corr} = {}^{9x}Zr_{Meas} - {}^{98}Mo_{Meas}\left[\left(\frac{^{9x}Mo}{^{98}Mo}\right)_{Ref}\left(\frac{M_{9x}}{M_{98}}\right)^{\beta}\right] - {}^{99}Ru_{Meas}\left[\left(\frac{^{9x}Ru}{^{99}Ru}\right)_{Ref}\left(\frac{M_{9x}}{M_{99}}\right)^{\beta}\right] \quad (1)$$

where '*Meas*' denotes measured beam intensities, '*Ref*' the reference natural Mo and Ru isotope ratio taken from IUPAC, $\beta$ the mass-fractionation coefficients, and $M_{9x}$ the atomic mass of the corresponding Zr, Mo or Ru isotope, also taken from IUPAC. While interference correction itself is important, the impact of a mass-bias correction during interference correction was found to be negligible (< 1 ppm offset on the internally normalized $^{96}$Zr/$^{90}$Zr values). The interference corrected intensities were then used to calculate Zr isotope ratios ($^{91}$Zr/$^{90}$Zr, $^{92}$Zr/$^{90}$Zr, $^{94}$Zr/$^{90}$Zr and $^{96}$Zr/$^{90}$Zr).

At this point the Zr isotope ratios are OPZ and interference corrected, but still have to be corrected for instrumental mass fractionation. As in the interference correction step, mass-bias correction was initially done by using the exponential law[52], where the measured ($R_{Meas}$) and true ($R_{Ref}$) ratio of two isotopes in the samples are related as function of the ratio of their atomic masses by a mass-fractionation coefficient $\beta$ as:

$$R_{Meas} = R_{Ref}\left(\frac{M_i}{M_j}\right)^{\beta} \quad (2)$$

with $M$ denoting atomic masses, and the subscripts $i$ and $j$ the numerator and denominator isotopes, respectively. Eq. 2 was used to infer $\beta$ from the measured and reference value for $^{94}$Zr/$^{90}$Zr, and the $\beta$ value thus obtained was used to correct the other $^x$Zr/$^{90}$Zr ratios. This approach was taken because the exponential law, which was originally derived empirically to describe the mass-fractionation on TIMS (thermal ionization mass-spectrometer) instruments[52,54–58], has been shown to adequately correct, to first-order, the mass-fractionation for MC-ICP-MS instruments[59–63].

As precision of MC-ICP-MS instruments rapidly improved over time, it became apparent that the exponential law does not perfectly describe mass-fractionation in these mass spectrometers[64–68]. When residual mass-dependent effects were observed in the internally normalized ratios obtained after correction with the exponential law (*Section 6.2*), a second approach was then tried, using the generalized power law (GPL). Introduced by Marechal et al. (1999), the GPL formulates mass-dependent fractionation relationships as:

$$R_{Meas} = R_{Ref}\, g^{(M_i^n - M_j^n)} \qquad (3)$$

where $g$ is a mass-fractionation factor (just like $\beta$ in the exponential law) and varying the exponent $n$ is equivalent to changing the mass-fractionation law. Most natural laws (*i.e.*, derived from first-principles) and empirical laws are simply special cases of the GPL. For instance, the GPL yields the equilibrium law for n = -1 and the power law for n = +1. The kinetic law (which is identical to the exponential law), is obtained in the limit where n → 0. Here, we used the GPL to find the law that best describes mass-fractionation on each instrument from each participating laboratory, and to assess whether apparent differences in the absolute composition of the candidate RM 8299 (and SRM 3169 lot #130920 and #071226) were simply artifacts of subtle differences in mass-fractionation laws relevant to different instruments.

To assess the homogeneity of the Zr isotope composition of the various iRM aliquots, the data after mass-bias correction was finally recast in 'mu' notation, as part-per-million (ppm) deviations relative to the isotopic composition of the bracketing standard:

$$\mu^{9x}Zr = \left( \frac{(^{9x}Zr/^{90}Zr)_{Sample}}{(^{9x}Zr/^{90}Zr)_{Standard}} - 1 \right) \times 10^6 \qquad (4)$$

Such a sample-standard bracketing approach allows for the correction of residual biases that cannot otherwise be accounted for, including both non-exponential mass-dependent effects and instrument-specific mass-independent effects. Laboratory 1 used the iRM aliquot #12 as bracketing standard, while laboratories 2 and 3 used SRM 3169. Uncertainties are reported as 95% confidence intervals (*i.e.*, 2SE) and calculated as $\frac{2 \times \sigma_{Standard}}{\sqrt{n}}$, where $2 \times \sigma_{Standard}$ is the external reproducibility of repeat measurements of the standard bracketed by itself (measured at the same concentration as the sample) within a session, and $n$ is the number of repeat analyses of the sample solution considered (see Table 3).

## 5. Results

Internally normalized data from all three laboratories show that the nine aliquots of candidate RM 8299 have Zr isotopic compositions identical to each other within error, confirming at high-precision (2SD of ±2.5 ppm for $\mu^{91}Zr$ and $\mu^{92}Zr$, ± 7.5 ppm for $\mu^{96}Zr$) the isotopic homogeneity of the RM stock solution curated by NIST (Fig. 2 and Table 3). As expected, this composition is also identical to that of SRM 3169 (lot #130920), the Zirconium Standard Solution from which NIST prepared the candidate RM 8299, as well as another lot of SRM 3169 (lot #071226). The same is true for the mass-dependent data ($\delta^{94/90}Zr$, see *Section 6.4*), which confirms the lack of mass-dependent fractionation between candidate RM8299 and SRM 3169 lot #130920 (-0.002 ± 0.006 ‰, 2SE, n = 18). The data on other pure Zr solutions reveal significant variability in mass-dependent isotope effects compared to candidate RM 8299, from -0.538 ± 0.005 ‰ (n=23) for Alfa Aesar lot 03-14247H to +0.291 ± 0.004 ‰ (n=29) for the AMES Zr solution. The long-term average of the PlasmaCal solution (IPGP-Zr) is -0.056 ± 0.002 ‰ (n=98). Geostandard values cover a much tighter range around candidate RM 8299, from -0.058 ± 0.003 ‰ (n=74) for AGV-2, to +0.058 ± 0.003 (n=59) for RGM-2.

Given the homogeneity of candidate RM 8299, the average absolute ratios measured in each laboratory – internally normalized to $^{94}Zr/^{90}Zr = 0.3381$ (Ref.[53]) – were calculated and compared to literature values[28,47,53,69,70] (Fig. 3 and Table 4). While older studies investigated either unspecified "natural Zr" standards or single-element solutions from various manufacturers (*e.g.*, Alfa-Aesar, SPEX), there is good agreement between the isotopic composition of these materials and that of the candidate RM 8299 / SRM 3169 (this work) and the ZIRC-1 reference material[47] distributed by the Canadian National Research Council (NRC). Inspection of the high-precision data, however, reveals differences in the internally normalized isotopic composition (i) between candidate RM 8299 / SRM 3169 and NRC ZIRC-1 (Fig. 3 and Table 4), and (ii) between different laboratories on the very same candidate RM 8299 solution (Figs. 3, 4 and Table 4). As discussed below (*Section 6.2*), these differences reflect a combination of mass-independent effects and non-exponential mass-dependent effects during mass-spectrometric analyses.

## 6. Discussion

### 6.1. iRM homogeneity

Of the key requirements that an iRM must fulfill (Table 1), the primary goal of this work was to address those pertaining to the homogeneity (#1) and composition (#3 and #4) of the NIST candidate RM 8299 and SRM 3169 (lot #130920, and #071226). As shown in Figure 2, independent analysis of nine aliquots of the candidate RM 8299 by three laboratories confirms at high-precision that the iRM stock solution is isotopically homogeneous (thereby addressing requirement #1). Such thorough inter-laboratory calibration is essential and demonstrates that Zr isotope data measured against the same standard in different laboratories using different analytical methods can be taken at face value and compared to one another, as long as matrix effects and interferences are appropriately addressed.

Importantly, the candidate RM 8299 composition was found to be identical to that of multiple lots (#130920 and #071226) of SRM 3169 – the Zr concentration reference material from NIST – which has been used extensively in radiogenic ($^{92}$Nb-$^{92}$Zr system) and nucleosynthetic Zr isotope studies. This is important because it means that the literature data reported relative to SRM 3169 lots #130920 and #071226 (Refs.[15–23]) are effectively already expressed on the candidate RM 8299 scale, and no conversion factor need be applied for comparison with future data. Similarly, lower precision data reported in earlier studies[13,24], can be directly compared to the RM 8299 scale because no differences have been detected at the > 30 ppm level between NIST SRM 3169 and the other Zr standards used in these studies[28]. In contrast, the high-precision data of Akram et al. (2105) and Akram and Schönbächler (2016), measured relative to Alfa-Aesar Zr standards, require a secondary normalization relative to terrestrial standard rock to account for mass-independent isotopic variations in the Zr standards that they used[23,26,27].

### 6.2. The problem with absolute ratios

*6.2.1. Variations in absolute ratios*

Whereas the standard-bracketed data from all three laboratories unambiguously demonstrate the isotopic homogeneity of the candidate RM 8299 (and SRM 3169 lots #130920 and #071226), inspection of the results in absolute ratio space (after internal normalization to $^{94}Zr/^{90}Zr$) reveals average compositions that differ slightly from one another (Fig. 3). The differences are relatively small (30-120 ppm for $^{91}Zr/^{90}Zr$, 6-64 ppm for $^{92}Zr/^{90}Zr$, and 35-221 ppm $^{96}Zr/^{90}Zr$), but significantly larger than the external reproducibility of the measurements (± 10-12 ppm on $^{91}Zr/^{90}Zr$ and $^{92}Zr/^{90}Zr$, and ±30 ppm on $^{96}Zr/^{90}Zr$). These differences could in principle be due to (i) improper OPZ and/or interference correction, (ii) non-exponential mass-dependent instrumental fractionation, (iii) mass-independent instrumental fractionation, or (iv) a combination thereof. Below, we examine each of these possibilities to elucidate which may be responsible for these differences in absolute ratios.

If these aforementioned differences in absolute ratios stemmed from improper OPZ and/or interference correction, then varying the magnitude of these corrections (or, in the most conservative scenario, turning them off entirely) would significantly affect the absolute ratios. Systematic investigation of these effects (see below) reveals that, while improper OPZ and interference correction slightly contribute to the differences in absolute ratios observed between laboratories, they are not the main control behind these differences.

OPZ correction and ArArO$^+$ interference. The OPZ beam intensities measured in the run acid reflect the low amounts of elements in the run acid itself, as well as residual elements leached from the sample introduction system surfaces. While variations in the OPZ values will directly affect the measured isotopic ratios, the impact on the internally normalized ratios will be negligible, as long as the background elements contributing to the OPZ (Zr, Mo, and Ru) have natural compositions. Comparison of the data treated with and without OPZ correction mainly confirms these expectations. Indeed, whereas the measured $^{91}Zr/^{90}Zr$ and $^{92}Zr/^{90}Zr$, $^{94}Zr/^{90}Zr$ and $^{96}Zr/^{90}Zr$ ratios of individual analyses calculated with and without OPZ correction differ by up to ±15, ±30, ±60 and ±150 ppm respectively, the same ratios after internal normalization are identical within ±4-6 ppm for $^{91}Zr/^{90}Zr$ and $^{92}Zr/^{90}Zr$. This clearly rules out an improper OPZ correction as the source of the differences seen between laboratories for $^{91}Zr/^{90}Zr$ and $^{92}Zr/^{90}Zr$. For $^{96}Zr/^{90}Zr$, residual positive anomalies subsist after internal normalization, ranging from 0 to +100 ppm. These $^{96}Zr$ positive anomalies most certainly are due to ArArO$^+$ interferences, as indicated by the elevated OPZ $^{96}Zr$ beam intensities compared to what would be expected based on the $^{90}Zr$ and Mo isotope intensities. Because the ArArO$^+$ interference will affect the OPZ and sample measurements similarly, its impact on the corrected ratios will be smaller than the value listed above (likely by an order of magnitude). Assuming the ArArO+ interference is stable within 10-25 % will result in a maximum variability of ~ ±10-25 ppm on the internally corrected $^{96}Zr/^{90}Zr$ ratios. This is small compared to the 221 ppm range observed in the average value measured in the different laboratories.

Mo and Ru isobaric interferences. If present, both molybdenum ($^{92}Mo$, $^{94}Mo$, and $^{96}Mo$) and ruthenium ($^{96}Ru$) will directly interfere on Zr isotopes. The high-purity of the candidate RM 8299 solutions, however, resulted in barely detectable ion beams on $^{98}Mo$, $^{99}Ru$ and $^{101}Ru$ (<10 µV after OPZ correction) as well as $^{100}Mo$ (<20 µV). As a result, Mo interference correction done using $^{98}Mo$ (Lab 1) or $^{100}Mo$ (Lab 2) had virtually no effect on the internally normalized $^{91}Zr/^{90}Zr$ and $^{92}$Zr/$^{90}$Zr (< 1 ppm), and changed the $^{96}$Zr/$^{90}$Zr by less than 6 ppm in Lab 1 and 10 ppm in Lab 2. Similarly, the Ru correction was found to be negligible, changing the internally normalized $^{96}$Zr/$^{90}$Zr of individual analyses by less than 1.5 ppm.

Interference from Zr hydrides (ZrH$^+$). Ion beam intensities were clearly observed at mass 93 (up to 140 µV) and mass 95 (up to 50 µV). Given the near absence of detectable Mo in the candidate RM 8299 solutions (*i.e.*, $^{98}$Mo < 10 µV in Lab 1, $^{100}$Mo < 20 µV in Lab 2), the signals on masses 93 and 95 must predominantly reflect the formation of hydrides: $^{92}$ZrH$^+$ and $^{94}$ZrH$^+$. The hydride formation rate, calculated as $^{9x}$ZrH$^+$/$^{9x}$Zr varied between ~1e-6 and 6e-6, indicating that hydride formation could shift the internally normalized $^{91}$Zr/$^{90}$Zr and $^{92}$Zr/$^{90}$Zr ratios by ~5-25 ppm and ~1-4 ppm, respectively. Like the other effects considered above, the impact of hydride formation is small and cannot, by itself, explain the difference in absolute ratios obtained in the different laboratories.

Mo interference correction using mass 95. For the internally normalized but non-sample standard bracketed ratios considered here, a potential complication arises if mass 95 is used to correct for Mo isotope interference (as could have been done in Lab 1 and 2, which both monitored mass 95). The signal at mass 95, being predominantly due to $^{94}$ZrH$^+$, would yield to overcorrection of Mo interferences. The impact of this effect is minor, resulting in shifts of less than 0.5, 2 and 15 ppm, respectively, on the internally normalized $^{91}$Zr/$^{90}$Zr, $^{92}$Zr/$^{90}$Zr and $^{96}$Zr/$^{90}$Zr ratios. These values are, however, 2-3 times larger than the interference corrections using Mo masses free of hydrides (see above). To avoid systematic biases due to hydride formation, we therefore decided to either correct Mo interferences using masses free of hydrides (*i.e.*, $^{98}$Mo in Lab 1 and $^{100}$Mo in Lab 2) or to apply no Mo correction (Lab 3, where no Mo masses were monitored). It is noteworthy that this systematic bias can be accounted for by intensity-matched sample-standard bracketing, and $^{95}$Mo is a suitable interference monitor for such measurements, most often used in the literature (*e.g.*, Refs.[17,19,21,23,25,27]).

Other spectral interferences. Two other spectral interferences are likely to be present during the measurements: ArArN$^+$ on mass 94 and ArSiSi$^+$ on mass 96. The first will affect the mass-bias correction, and for all laboratories, will result in shifts of the three isotope ratios along mass-fractionation lines (see next section). The second would manifest as a mass-independent isotope effect leading to elevated $^{96}$Zr/$^{90}$Zr ratios. The production of such an ArSiSi$^+$ interference will mainly occur in Labs 1 & 3, as Si from the glass spray chamber might be leached by the small amounts of HF present in the run acid. The data, however, shows that the $^{96}$Zr/$^{90}$Zr measured in both Labs1 & 3 is lower than that measured in Lab 2, suggesting a minor contribution from ArSiSi+ interferences.

Since the impact of OPZ and interference corrections appears to be somewhat limited and unlikely to fully explain the differences in absolute ratios measured in the different laboratories, we now assess the impact of the nature of the instrumental mass-bias on the data. To do so, we plot the internally-normalized ratios as a function of the mass-fractionation coefficient, $\beta$, calculated using the exponential law (Eq. 2) (Fig. 4). In this space, specific predictions can be made about inter- and intra-laboratory data distribution based on the origin of the variability.

- *Mass-independent effects:* In the hypothetical scenario where the instrumental mass-bias is *perfectly* described by the exponential law (and without interferences on the normalizing isotopes), the internal normalization procedure would correct all mass-dependent fractionation. In this case, the ratios measured in each laboratory would be independent of the magnitude of the instrumental mass-bias and the observed differences in absolute ratios obtained in each laboratory would reflect mass-independent effects. In an isotope ratio vs. $\beta$ plot, the data from any given laboratory would therefore plot along a unique horizontal line (*i.e.*, isotope ratios are independent from $\beta$), and the data from all three laboratories would define three distinct lines.

- *Non-exponential mass-dependent effects:* In the absence of mass-independent effects, the different average absolute ratios obtained in each laboratory might reflect the fact that instrumental mass-bias is improperly described by the exponential law. In this case, even after internal normalization using the exponential law, a residual (mass-dependent) component of the instrumental mass-bias would still affect the data. The different ratios obtained in the different labs would thus reflect how much the mass-fractionation law on each instrument deviates from the exponential law. Graphically, this would result in covariations between the internally normalized ratios and the $\beta$ values calculated using the exponential law. Note that an interference affecting one, or both, of the normalizing isotopes would manifest in the same way.

As shown in Figure 4, both mechanisms seem to be influencing the data we report. Within a given laboratory, subtle to very pronounced trends are observed between internally normalized ratios and $\beta$ values, demonstrating that mass-dependent fractionation was not fully corrected by the exponential law. At the same time, for a given mass-bias coefficient, differences in absolute ratios are observed between laboratories (see Lab 1 and 2, at $\beta \sim 1.725$), and between analytical sessions within the same laboratory (Lab 3, at $\beta \sim 2.25$), indicating that mass-independent effects are affecting the data.

*6.2.2. Mass-fractionation law versus mass-independent effects*

To gain further insights into these effects, it is most useful to compare the internally normalized ratios (denoted as $R^*$) in log-log triple isotope plots: *i.e.*, $\text{Ln}(R^*_{4/1})$ versus $\text{Ln}(R^*_{3/1})$, where the subscripts describe any isotope ratios $i_4/i_1$ and $i_3/i_1$, both normalized to isotope ratio $i_2/i_1$. Indeed, in such a space, non-exponential mass-dependent effects will result in correlations whose slopes, *s*, depend on the nature of the mass-fractionation law, as (see Appendix for derivation):

$$s^{Ln(R^*_{4/1})}_{Ln(R^*_{3/1})} = \frac{\theta^n_{4/1-2/1} - \theta^k_{4/1-2/1}}{\theta^n_{3/1-2/1} - \theta^k_{3/1-2/1}}, \quad (5)$$

where *k* represents the GPL exponent for the law used for internal normalization and *n* the exponent for the law actually describing the data, and the mass fractionation exponent, $\theta^n_{x/1-2/1}$, is defined as:

$$\theta^n_{x/1-2/1} = (M^n_x - M^n_1)/(M^n_2 - M^n_1). \quad (6)$$

In equation (6) above, the isotope ratio $i_2/i_1$ is used for internal normalization, and *x* denotes a third isotope used to calculate the ratio $i_x/i_1$. When using the exponential law, equation (6) takes the form:

$$\theta^{Exponential}_{x/1-2/1} = Ln(M_x/M_1)/Ln(M_2/M_1). \qquad (7)$$

In contrast, mass-independent effects (*e.g.*, variable cup efficiencies, counting fluctuations) will appear as deviations from these expected trends. For instance, and as thoroughly discussed in Ref.[67], fluctuations in cup efficiency result in changes in the intercept of the mass-fractionation curves (their Eq. 48), while noise due to counting fluctuations will result in trends with slopes distinct from those of the mass-fractionation curves (their Eq. 63).

As can be seen in Figure 5, and as suggested by Figure 4, the intra- and inter-laboratory variability in absolute ratios results from a combination of both non-exponential mass-dependent fractionation and mass-independent effects. Notably, Figure 5 reveals that the offsets between data from Lab 1 (orange symbols) and Lab 2 (grey symbols) predominantly stem from different magnitudes of non-exponential mass-dependent fractionation during analysis, while those between data from Labs 1-2 and Lab 3 (blue symbols) are mainly due to instrument-specific mass-independent effects. The nature of the law (*i.e.*, the value of *n* for the GPL) on each instrument can be determined based on the slope defined by the raw ratios in log-log space (see Eq. (S3) in Appendix). This calculation (excluding ratios involving $^{91}$Zr, see below) reveals that in Labs 1 and 2 (Session 1), the data is best explained by the GPL using an exponent *n* of ~1 (*i.e.*, power law), while for Lab 3, a lower exponent is obtained (n ~ 0.1-0.5), consistent with the fact that the exponential law in Lab 3 results in mostly uncorrelated isotope variations (*i.e.*, mass-dependent fractionation was adequately corrected) (Figure 5, lower panel). Examination of intra-laboratory trends reveals further nuances and indicates a contribution from mass-independent counting noise (grey arrows), primarily affecting $^{91}$Zr, consistent with the twice larger variations (per unit mass) seen in internally normalized $^{91}$Zr/$^{90}$Zr compared to the other ratios (Figure 4). This is reminiscent of the minor interference observed on $^{91}$Zr by Ref.[28], and as noted in Section 6.2.1 is most likely due to $^{90}$Zr hydrides ($^{90}$ZrH$^+$) interfering on $^{91}$Zr. Finally, the clear session-to-session offsets seen in a given laboratory, and following neither the non-exponential mass-bias trend nor the counting noise trends (this is particularly pronounced in the $^{96}$Zr/$^{90}$Zr vs $^{92}$Zr/$^{90}$Zr data for Lab 3) indicates that other mass-independent effects (red arrows) are affecting the data. These could stem from session-to-session variations in cup efficiencies or tuning of the sample inlet system.

The identification of these lab- and instrument-specific effects was only made possible by the careful approach to inter-laboratory calibration performed here, and demonstrates the difficulty of determining the absolute composition of iRMs at high-precision. In fact, when working at the < 10 ppm level of precision, the *accurate* determination of absolute ratios using MC-ICP-MS might simply be unattainable as (i) the nature of instrumental mass-fractionations in these instruments remain not well-understood and cannot at present be fully quantified, and (ii) a flurry of molecular interferences generated by the ionization in the Ar plasma and subsequent extraction of the ions into a high vacuum zone can affect the data. In light of these findings, the recommended high-precision absolute ratios calibrated in only one laboratory cannot be applied to another laboratory, and the only way to ensure data consistency is for all laboratories to use the very same standard to bracket sample measurements and report all data as deviation relative to the standard (*e.g.*, $\mu^{9x}$Zr, Eq. (4)), even if the composition of the latter is imperfectly known.

### *6.3. Candidate RM 8299 as an ideal community standard*

To ensure direct comparability of data produced in different laboratories, it is imperative that all data are measured (or at the very least, reported) against the exact same standard. Because of the rapid expansion of the Zr isotope system, multiple Zr isotope standards have been proposed, in particular for the characterization of mass-dependent effects, such as the SRM 3169 (Refs.[26,29–35]), the candidate RM 8299 presented herein[7–9,11,12,36–38], a single-element Zr solution from PlasmaCal (SCP Science) also known as the IPGP-Zr standard[5,6,10,11,30,33,35,38–46], and the NRC-ZIRC 1 (Refs.[44,47,48]). When the mass-independent literature is included, this list expands to include a single-element Zr solution from Johnson Matthey[13], a standard solution prepared from AMES Zr metal[14], at least two lots of NIST SRM 3169 (Refs.[13,15–23]), as well as 4 different lots of single-element Zr solutions from Alfa-Aesar[19,24–27]. Faced with such a proliferation of standards, it is valid to ask whether any of these materials are a suitable primary iRM against which all isotope data, both mass-dependent effects and isotope anomalies, should be reported.

In Table 5, we consider how the various previously utilized RMs meet (or fall short of) the community-identified criteria for an iRM. Since all proposed RMs are purified Zr solutions, they all fulfill the requirements of homogeneity (#1) and purity (#2). Not all proposed RMs, however, have compositions representative of a major geological reservoir (#3), as significant mass-dependent variations exist between these standards (see Section 6.4). While the IPGP-Zr solution displays a $\delta^{94/90}$Zr value close to that of Ocean Island Basalts (OIBs) (*i.e.*, ~ 0.048 ± 0.032 ‰ lighter than OIBs, Ref.[6]), only the NIST solutions have $\delta^{94/90}$Zr value identical to that of OIBs (~ -0.008 ± 0.032 ‰). Relative to the NIST SRM 3169 lots used in this study (and therefore candidate RM 8299, which has identical isotopic composition), isotope anomalies (#4) have been documented in three lots of Alfa-Aesar Zr solutions[23,26,27]. Similar effects have been observed for other elements (*e.g.*, Ti[71], Ni[72], Mo[73]; W[74]), and have been interpreted as possibly stemming from the elemental purification/enrichment production process (*e.g.*, Kroll process) and/or non-exponential mass-dependent fractionation during the genesis of the ores used to make these standards. While no data are currently available for IPGP-Zr or NRC ZIRC-1, it is reasonable to expect small degrees of mass-independent isotope variations in them as a result. In contrast, high-precision data on terrestrial geostandards[23] show no isotope anomalies relative to NIST SRM 3169 lot #130920 (and thus candidate RM 8299) within ±2 ppm for $\mu^{91}$Zr and $\mu^{92}$Zr, and within ±4-7 ppm for $\mu^{96}$Zr.

Beyond the composition of an ideal iRM, its curation and distribution are also key considerations. NIST RM 8299 (and thus the SRM 3169 lots #130920 and #071226) is the only material that the community has come together to establish as the iRM for Zr isotopes (#5a in Table 5), and it is by far the most used standard against which mass-independent and mass-dependent data have been reported with a total of 15 publications: nearly as many as all other proposed RMs combined (#5b). A conflict-of-interest free distribution (#6) is only ensured for the NIST and NRC standards because other proposed RMs are owned by individual research groups, rather than neutral institutions dedicated to the development, curation, and distribution of RMs. Similarly, only the NIST and NRC standards have been produced in sufficient stock to last decades (#7).

Of all proposed RMs, it is thus clear that *only the NIST candidate RM 8299 fulfills all requirements for an ideal iRM*, and as such, we recommend that candidate RM 8299 be considered the scale defining iRM for Zr isotopes. This will have multiple advantages for the Zr isotope community. First, reporting of Zr isotope values against candidate RM 8299 will enable direct comparison of future data produced in different laboratories. Second, since SRM 3169 lots #130920 and #071226 are isotopically indistinguishable from candidate RM 8299, this will ensure direct comparison of newly produced data with the vast majority of existing literature on mass-independent isotopes data. Third, having a unique iRM used for all Zr isotope investigations will make it possible for isotope anomaly data obtained in one laboratory to be used to correct mass-dependent effects measured in another laboratory.

For those laboratories using a different Zr standard, all data should be made traceable to candidate RM 8299 through careful cross-calibration to determine the presence of any isotopic anomaly and mass-dependent fractionation relative to candidate RM 8299. This is particularly important as clear isotope anomalies and/or mass-dependent fractionation have already been observed in virtually all Zr standards used in the literature. At the current level of precision, only the SRM 3169 lots #130920 and #071226 have been found to have identical isotopic composition to candidate RM 8299. We note that SRM 3169 lot #130920 is currently out of stock and a new standard is in production, which may use a different Zr metal source compared to previous SRM 3169 lots. As such, it is very likely that this next lot of Zr SRM 3169 will have a different Zr isotope composition, and as such it is not suggested that this upcoming lot be used as an isotopic reference material.

### *6.4 Reference materials $\delta^{94/90}Zr_{RM\ 8299}$ conversion factors*

To facilitate conversion of literature data reported against other materials to the RM 8299 scale, Table 6 presents a summary of $\delta^{94/90}Zr_{RM-8299}$ values for the main Zr RMs used in the literature. With the exception of the GJ-1 zircon and NRC ZIRC-1 material, whose values were calculated based on their difference with the IPGP-Zr standard as reported in Refs.[42,44], we determined all other conversion factors through repeat analyses over several years. Figure 6 shows the details of the data used to calculate the average conversion factors presented in Table 6.

Data reported against another Zr standard (*e.g.*, ZIRC-1, IPGP-Zr), can easily be converted to the RM 8299 scale by adding the $\delta^{94/90}Zr$ value of the standard reported in Table 6. For instance, conversion of data reported against ZIRC-1, are converted to the RM 8299 scale as:

$$\delta^{94/90}Zr_{RM\ 8299} = \delta^{94/90}Zr_{ZIRC-1} + (-0.276 \pm 0.006) \qquad (8),$$

while conversion of data reported against IPGP-ZR can be done as:

$$\delta^{94/90}Zr_{RM\ 8299} = \delta^{94/90}Zr_{IPGP-Zr} + (-0.056 \pm 0.002) \qquad (9).$$

Data for several widely used geostandards (andesite AGV-2, basalts BCR-2 and BHVO-2, and rhyolite RGM-2) are also reported in Table 6. As for the pure Zr solutions, the long-term average values and their associated uncertainties were determined by repeat measurements carried out over more than 3 years (see dates on top of Fig. 6).

### 7. Conclusion

The field of Zr stable isotopes sorely needs an isotope reference material (iRM) that adheres to community-defined best-practices (Table 1). RM 8299, was thus developed in partnership with NIST, and calibrated by the Zr isotope community (Fig. 1). Our data show that 9 independently bottled aliquots of RM 8299, as well as 2 different lots of SRM 3169, have identical compositions within the resolution of modern instrumentation, which is ±2.5 ppm for $^{91}$Zr/$^{90}$Zr and $^{92}$Zr/$^{90}$Zr ratios, and ±7.5 ppm for the $^{96}$Zr/$^{90}$Zr ratio (Fig. 2, Table 3 & 4). Compared to other proposed reference materials, only NIST RM 8299 meets all requirements that an ideal iRM should fulfill (Table 5) and we therefore recommend that all future data (radiogenic, nucleosynthetic and mass-dependent) be reported against RM 8299. This will allow for direct comparison of all newly produced data (mass-dependent and independent) with the vast majority of existing literature data on isotope anomalies. To facilitate reporting of mass-dependent effects on the RM 8299 scale, high-precision conversion factors between other Zr standards and RM 8299 have been measured and/or compiled (Table 6 and Fig. 6).

Careful examination of mass-bias corrected (but not sample-standard bracketed) absolute ratios revealed differences (i) from session to session in any given laboratory, and (ii) between laboratories (Figs. 3 and 4). These intra- and inter-laboratory differences are best explained as a combination of non-exponential mass-dependent fractionation and instrument-specific mass-independent effects (*e.g.*, possibly related to cup degradation, or sample inlet system tuning) (Fig. 5). These results highlight the challenge associated with the *accurate* determination of absolute ratios using MC-ICP-MS in the absence of a more robust fundamental understanding of instrumental mass-fractionation in these instruments. As our inter-laboratory calibration shows, however, perfect knowledge of the absolute composition of the iRM is not necessary to ensure data comparability at high-precision, so long as all laboratories use the very same standard to bracket sample measurements and report the data.

**Appendix**

Correction of instrumental mass-bias using a law (*e.g.*, the exponential law) that does not perfectly describe the true instrumental mass-bias will result in correlations between corrected ratios (*e.g.*, Fig. 5). Here we provide the details of the derivation of Eq. (5), which predicts the slope of the correlation between two corrected ratios. We note that similar treatments exist in the literature (*e.g.*, Refs.[67,75–79]), but we find it useful to revisit the question here for the sake of clarity, in particular given the variable notations used in previous work.

Let us consider four isotopes, $i_1$, $i_2$, $i_3$ and $i_4$, of respective mass $M_1$, $M_2$, $M_3$ and $M_4$, and denote $R_{i/j}$ the ratio of any two isotopes of mass $M_i$ and $M_j$. Following the exponential law[52] to correct the mass-fractionation experienced by the sample (in nature and in the instrument), the measured and 'true' ratios are related to their atomic masses through a mass-fractionation factor, $\beta$ (sometimes also denoted *f*, *e.g.*, Ref.[59]), as:

$$R_{i/j}^{meas} = R_{i/j}^{true} \left(\frac{M_i}{M_j}\right)^\beta \quad (S1).$$

As shown by Ref.[61], the exponential law is a special case of a generalized power law (GPL), which relates the measured and 'true' ratio as:

$$R_{i/j}^{meas} = R_{i/j}^{true} \, g^{(M_i^n - M_j^n)} \quad (S2)$$

where *g* is a mass-fractionation factor and *n* is a free-parameter that determines the mass-fractionation law. In the limit where n → 0, the GPL reduces to the exponential law.

Provided that the sample mass-fractionation follows the general form of the GPL, important insights into the nature of the mass-fractionation law (*i.e.*, the value of *n*) can be obtained by plotting two measured ratios against one another in log-log space. Indeed, in such spaces, linear alignments should be observed whose slopes, *s*, depend solely on the mass of the isotopes considered and the value of the exponent *n* (Eq. 16 in Ref.[67]), as:

$$s_{Ln(R_{2/1})}^{Ln(R_{3/1})} = \frac{Ln\left(R_{3/1}^{meas}/R_{3/1}^{true}\right)}{Ln\left(R_{2/1}^{meas}/R_{2/1}^{true}\right)} = \frac{(M_3^n - M_1^n)}{(M_2^n - M_1^n)} \quad (S3).$$

The term of the right-hand side (RHS) of Eq. (S3) is known by many names in the literature. In Ref.[67], it is identified as the slope between two measured ratios and denoted $s_{j/k}^{i/k}$. In Ref.[80] is it referred to as the exponent $\beta$ (their eqs 15, 21 and 25), and is defined as the exponent relating the fractionation factors, $\alpha^{a-b}$, of two isotope ratios in two substances *a* and *b*, as $\alpha_{2/1}^{a-b} = \left(\alpha_{3/1}^{a-b}\right)^\beta$. Similarly, it is called $\beta^{\text{inferred}}$ in Ref.[66] and $\beta^{\text{law}}$ in Ref.[79], where the superscript terms 'inferred' and 'law' stand in for the *n* exponent of the GPL (Eq. 2 in both papers). In Ref.[78] it is referred to as the exponent $\beta^{i3}$ (their Eq. 1). To avoid confusion with the mass-fractionation factor $\beta$ of the exponential law, Ref.[76] denoted this exponent as $^{i1,i2,i3}\theta$ (their Eq. 3). Here, we follow the spirit of Ref.[76] and define the mass fractionation exponent, $\theta_{3/1-2/1}^n$, as:

$$\theta_{3/1-2/1}^n = (M_3^n - M_1^n)/(M_2^n - M_1^n) \quad (S4)$$

where the exponent *n* is the same exponent as in the GPL, and the subscript denotes the isotope ratios considered. When using the exponential law, equation (S4) takes the form:

$$\theta_{3/1-2/1}^{Exponential} = Ln(M_3/M_1)/Ln(M_2/M_1). \quad (S5)$$

Having established this nomenclature, we are finally ready to consider how fractionation-corrected ratios will correlate if the mass-fractionation law used for correction is different from the true mass-fractionation law inside the instrument. To correct the instrumental mass-bias, it is customary to fix an isotopic ratio, say $i_2/i_1$, to an assumed 'true' value, in order to calculated the mass-fractionation factor as:

$$Ln(g) = Ln\left(R_{2/1}^{meas}/R_{2/1}^{true}\right)/(M_2{}^n - M_1{}^n) \quad (S6),$$

The RHS of Eq (S6) can then be substituted into Eq (S2), to correct a second ratio, say $i_3/i_1$, as:

$$Ln\left(R_{3/1}^{corr}\right) = Ln\left(R_{3/1}^{meas}\right) - (M_3{}^n - M_1{}^n) Ln(g)$$

$$= Ln\left(R_{3/1}^{meas}\right) - \frac{(M_3{}^n - M_1{}^n)}{(M_2{}^n - M_1{}^n)} Ln\left(\frac{R_{2/1}^{meas}}{R_{2/1}^{true}}\right) \quad (S7).$$

which can simply be rewritten as:

$$Ln\left(R_{3/1}^{corr}\right) = Ln\left(R_{3/1}^{meas}\right) - \theta_{3/1-2/1}^n Ln\left(\frac{R_{2/1}^{meas}}{R_{2/1}^{true}}\right) \quad (S8).$$

Following Ref.[75], if the wrong law (exponent $k$ instead of $n$) was used to correct the data, the inappropriately corrected ratio $R^*$ would be calculated as:

$$Ln\left(R^{*corr}_{3/1}\right) = Ln\left(R_{3/1}^{meas}\right) - \theta_{3/1-2/1}^k Ln\left(\frac{R_{2/1}^{meas}}{R_{2/1}^{true}}\right) \quad (S9).$$

The difference between the inappropriately corrected ratio and the true ratio, is then calculated by subtracting Eq (S8) from (S9):

$$Ln\left(R^{*corr}_{3/1}\right) - Ln\left(R_{3/1}^{corr}\right) = \left(\theta_{3/1-2/1}^n - \theta_{3/1-2/1}^k\right) Ln\left(\frac{R_{2/1}^{meas}}{R_{2/1}^{true}}\right) \quad (S10).$$

The above equation is notable because it shows that the magnitude of departure of the corrected ratio from the true ratio, $\varepsilon_{3/1} = \left(\frac{R^{*corr}_{3/1}}{R_{3/1}^{corr}} - 1\right) 10^4$, directly depends on the extent of mass-fractionation. Indeed, using (i) the δ' notation, where δ'= $Ln\left(R_{i/j}/R_{i/j}^{reference}\right) 10^3$, to quantify the mass-fractionation, and (ii) the fact that $Ln(x) \approx (x-1)$ when x is close to unity, Eq. (S10) can be rewritten in terms of isotope anomaly as (Eq. 12 in Ref.[77]):

$$\varepsilon_{3/1} = 10 \left(\theta_{3/1-2/1}^n - \theta_{3/1-2/1}^k\right) \delta'_{2/1} \quad (S11).$$

Now considering a second corrected ratio, $i_4/i_1$, an equation similar to (S10) can be written:

$$Ln\left(R^{*corr}_{4/1}\right) - Ln\left(R_{4/1}^{corr}\right) = \left(\theta_{4/1-2/1}^n - \theta_{4/1-2/1}^k\right) Ln\left(\frac{R_{2/1}^{meas}}{R_{2/1}^{true}}\right) \quad (S12).$$

Substituting Equation Eq (S10) into (S12), yields:

$$\boldsymbol{Ln\left(R^{*corr}_{4/1}\right) - Ln\left(R_{4/1}^{corr}\right) = \frac{\left(\theta_{4/1-2/1}^n - \theta_{4/1-2/1}^k\right)}{\left(\theta_{3/1-2/1}^n - \theta_{3/1-2/1}^k\right)} \left[Ln\left(R^{*corr}_{3/1}\right) - Ln\left(R_{3/1}^{corr}\right)\right]} \quad (S13).$$

Equation (S13) indicates that if the wrong law (exponent *k* instead of *n*) is used to correct the data, a residual correlation is expected between corrected-ratios, and the slope of this correlation is:

$$S_{Ln(R^*_{3/1})}^{Ln(R^*_{4/1})} = \frac{\theta^n_{4/1-2/1} - \theta^k_{4/1-2/1}}{\theta^n_{3/1-2/1} - \theta^k_{3/1-2/1}}, \quad (S14).$$

Equation (S14) can be rewritten in terms of isotope anomalies (ε), to clarify that mass-bias correction using an inappropriate mass-fractionation law will result in correlated isotope anomalies between various isotopes:

$$\varepsilon_{4/1} = \frac{\left(\theta^n_{4/1-2/1} - \theta^k_{4/1-2/1}\right)}{\left(\theta^n_{3/1-2/1} - \theta^k_{3/1-2/1}\right)} \varepsilon_{3/1} \quad (S15).$$

Finally, we note that an approximate form of Eq (S15) was derived by Ref.[75] using Taylor series expansion, as (their equation A14, as well as Eq 35 in Ref.[76]):

$$\varepsilon_{4/1} = \frac{(M_4 - M_1)(M_4 - M_2)}{(M_3 - M_1)(M_3 - M_2)} \varepsilon_{3/1} \quad (S16).$$

This approximation is slightly more practical than Eq (S15), but it is important to realize that the exponents *k* and *n* (which describe the mass-fractionation used and the true mass-fractionation) no longer appear in Eq (S16), and as such, no insights into the nature of the mass-fractionation laws relevant to the measurement can be gained when using this equation.

**Author contributions:**
**François L.H. Tissot:** Conceptualization, Funding acquisition, Project administration, Methodology, Investigation, Formal analysis, Resources, Validation, Visualization, Writing – Original draft, Data Curation. **Mauricio Ibañez-Mejia:** Conceptualization, Funding acquisition, Project administration, Methodology, Investigation, Resources, Validation, Writing – Review & Editing. **Savelas A. Rabb:** Methodology, Resources, Investigation. **Rebecca A. Kraft:** Resources, Validation, Writing – Review & Editing. **Robert D. Vocke:** Methodology, Resources, Investigation. **Manuela A. Fehr:** Methodology, Investigation, Validation, Writing – Review & Editing. **Maria Schönbächler:** Methodology, Resources, Validation, Writing – Review & Editing. **Haolan Tang:** Methodology, Investigation, Writing – Review & Editing. **Edward D. Young:** Resources, Writing – Review & Editing.

**Conflict of interest:** Authors declare no competing interests.

**Acknowledgments:** We thank Jan Render for providing aliquots of the AMES Zr and Alfa-Aesar Zr solutions. **Funding:** This work was supported by NSF-EAR grants 1824002 (to FLHT) and 2131632 and 2143168 (to MIM) and start-up funds to FLHT provided by Caltech. MS acknowledges funding of the Swiss National Science Foundation (project 200021_208079). **Data and materials availability:** All data needed to evaluate the conclusions in the paper are present in the paper and/or the Supplementary Materials. Additional data related to this paper may be requested from the authors, and additional samples can be procured from the Smithsonian Institution.

**References:**

1  J. W. Valley, D. R. Cole and Mineralogical Society of America, Eds., *Stable isotope geochemistry*, Mineralogical Society of America, 2001.
2  C. M. Johnson, B. L. Beard and F. Albarède, Eds., *Geochemistry of non-traditional stable isotopes*, Mineralogical Society of America ; Geochemical Society, 2004.
3  F.-Z. Teng, J. Watkins, N. Dauphas and Mineralogical Society of America, Eds., *Non-traditional stable isotopes*, Mineralogical Society of America ; Geochemical Society ; De Gruyter, [2017], 2017.
4  M. Ibañez-Mejia and F. L. H. Tissot, *Elements*, 2021, **17**, 379–382.
5  E. C. Inglis, J. B. Creech, Z. Deng and F. Moynier, *Chemical Geology*, 2018, **493**, 544–552.
6  E. C. Inglis, F. Moynier, J. Creech, Z. Deng, J. M. D. Day, F.-Z. Teng, M. Bizzarro, M. Jackson and P. Savage, *Geochimica et Cosmochimica Acta*, 2019, **250**, 311–323.
7  M. Ibañez-Mejia and F. L. H. Tissot, *Sci. Adv.*, 2019, **5**, eaax8648.
8  H. G. D. Tompkins, L. J. Zieman, M. Ibañez-Mejia and F. L. H. Tissot, *J. Anal. At. Spectrom.*, 2020, **35**, 1167–1186.
9  H. G. D. Tompkins, M. Ibañez-Mejia, F. L. H. Tissot, E. Bloch, Y. Wang and D. Trail, *Geochem. Persp. Let.*, 2023, **25**, 25–29.
10 J.-L. Guo, Z. Wang, W. Zhang, F. Moynier, D. Cui, Z. Hu and M. N. Ducea, *Proc Natl Acad Sci USA*, 2020, **117**, 21125–21131.
11 J.-L. Guo, Z. Wang, W. Zhang, L. Feng, F. Moynier, Z. Hu, L. Zhou and Y. Liu, *Earth-Science Reviews*, 2023, **237**, 104289.
12 H. M. Kirkpatrick, T. Mark Harrison, M. Ibañez-Mejia, F. L. H. Tissot, S. A. MacLennan and E. A. Bell, *Geochimica et Cosmochimica Acta*, 2023, S001670372300217X.
13 C. Sanloup, J. Blichert-Toft, P. Télouk, P. Gillet and F. Albarède, *Earth and Planetary Science Letters*, 2000, **184**, 75–81.
14 C. Münker, S. Weyer, K. Mezger, M. Rehkämper, F. Wombacher and A. Bischoff, *Science*, 2000, **289**, 1538–1542.
15 M. Schönbächler, M. Rehkämper, A. N. Halliday, D.-C. Lee, M. Bourot-Denise, B. Zanda, B. Hattendorf and D. Günther, *Science*, 2002, **295**, 1705–1708.
16 M. Schönbächler, D.-C. Lee, M. Rehkämper, A. N. Halliday, M. A. Fehr, B. Hattendorf and D. Günther, *Earth and Planetary Science Letters*, 2003, **216**, 467–481.
17 M. Schönbächler, D.-C. Lee, M. Rehkämper, A. N. Halliday, B. Hattendorf and D. Günther, *Geochimica et Cosmochimica Acta*, 2005, **69**, 775–785.
18 M. Schönbächler, M. Rehkämper, M. A. Fehr, A. N. Halliday, B. Hattendorf and D. Günther, *Geochimica et Cosmochimica Acta*, 2005, **69**, 5113–5122.
19 T. Iizuka, Y.-J. Lai, W. Akram, Y. Amelin and M. Schönbächler, *Earth and Planetary Science Letters*, 2016, **439**, 172–181.
20 J. Render and G. A. Brennecka, *Earth and Planetary Science Letters*, 2021, **555**, 116705.
21 M. K. Haba, Y.-J. Lai, J.-F. Wotzlaw, A. Yamaguchi, M. Lugaro and M. Schönbächler, *Proc Natl Acad Sci USA*, 2021, **118**, e2017750118.
22 C. Burkhardt, F. Spitzer, A. Morbidelli, G. Budde, J. H. Render, T. S. Kruijer and T. Kleine, *Sci. Adv.*, 2021, **7**, eabj7601.
23 J. Render, G. A. Brennecka, C. Burkhardt and T. Kleine, *Earth and Planetary Science Letters*, 2022, **595**, 117748.
24 W. Akram, M. Schönbächler, P. Sprung and N. Vogel, *ApJ*, 2013, **777**, 169.

25 W. Akram, M. Schönbächler, S. Bisterzo and R. Gallino, *Geochimica et Cosmochimica Acta*, 2015, **165**, 484–500.
26 W. Akram and M. Schönbächler, *Earth and Planetary Science Letters*, 2016, **449**, 302–310.
27 B.-M. Elfers, P. Sprung, N. Messling and C. Münker, *Geochimica et Cosmochimica Acta*, 2020, **270**, 475–491.
28 M. Schönbächler, M. Rehkämper, D.-C. Lee and A. N. Halliday, *Analyst*, 2004, **129**, 32–37.
29 L. Feng, W. Hu, Y. Jiao, L. Zhou, W. Zhang, Z. Hu and Y. Liu, *J. Anal. At. Spectrom.*, 2020, **35**, 736–745.
30 C. Huang, H. Wang, H.-M. Yu, L.-P. Feng, L.-W. Xie, Y.-H. Yang, S.-T. Wu, L. Xu and J.-H. Yang, *J. Anal. At. Spectrom.*, 2021, 10.1039.D1JA00160D.
31 Z. Hu, X.-H. Li, T. Luo, W. Zhang, J. Crowley, Q. Li, X. Ling, C. Yang, Y. Li, L. Feng, X. Xia, S.-B. Zhang, Z. Wang, J. Guo, L. Xu, J. Lin, X. Liu, Z. Bao, Y. Liu, K. Zong, W. Chen and S. Hu, *J. Anal. At. Spectrom.*, 2021, 10.1039.D1JA00311A.
32 P. Hu, T. Luo, J. Crowley, Y. Wu, C. Zhang, X. Xia, T. Long, S. Zhang, Z. Bao, L. Xu, L. Feng, W. Zhang and Z. Hu, *Geostandard Geoanalytic Res*, 2023, ggr.12495.
33 W. Zhang, Z. Hu, L. Feng, Z. Wang, Y. Liu, Y. Feng and H. Liu, *J. Earth Sci.*, 2022, **33**, 67–75.
34 Y. Jiao, L. Zhou, T. J. Algeo, J. Shen, L. Feng, Y. Hu, J. Liu, L. Chi and M. Shi, *Chemical Geology*, 2022, **610**, 121074.
35 Z. Zhu, W. Zhang, J. Wang, Z. Wang, J.-L. Guo, J. Elis Hoffmann, L. Feng, T. Luo, Z. Hu, Y. Liu and F. Moynier, *Geochimica et Cosmochimica Acta*, 2023, **342**, 15–30.
36 M. Méheut, M. Ibañez-Mejia and F. L. H. Tissot, *Geochimica et Cosmochimica Acta*, 2021, **292**, 217–234.
37 M. Klaver, S. A. MacLennan, M. Ibañez-Mejia, F. L. H. Tissot, P. Z. Vroon and M.-A. Millet, *Geochimica et Cosmochimica Acta*, 2021, **310**, 221–239.
38 J.-C. Xie, D.-C. Zhu, Q. Wang and L.-L. Zhang, *International Journal of Mass Spectrometry*, 2023, **484**, 116995.
39 W. Zhang, Z. Wang, F. Moynier, E. Inglis, S. Tian, M. Li, Y. Liu and Z. Hu, *J. Anal. At. Spectrom.*, 2019, **34**, 1800–1809.
40 T. Wu, W. Zhang and S. A. Wilde, *Chemical Geology*, 2020, **551**, 119776.
41 S. Y. Tian, F. Moynier, E. C. Inglis, J. Creech, M. Bizzarro, J. Siebert, J. M. D. Day and I. S. Puchtel, *Geochem. Persp. Let.*, 2020, **15**, 40–43.
42 S. Y. Tian, E. C. Inglis, J. B. Creech, W. Zhang, Z. Wang, Z. Hu, Y. Liu and F. Moynier, *Chemical Geology*, 2020, **555**, 119791.
43 S. Y. Tian, F. Moynier, E. C. Inglis, R. L. Rudnick, F. Huang, C. Chauvel, J. B. Creech, R. M. Gaschnig, Z. Wang and J.-L. Guo, *Earth and Planetary Science Letters*, 2021, **572**, 117086.
44 S. Y. Tian, F. Moynier, E. C. Inglis, N. K. Jensen, Z. Deng, M. Schiller and M. Bizzarro, *J. Anal. At. Spectrom.*, 2022, **37**, 656–662.
45 S. He, Y. Li, L.-G. Wu, D.-F. Guo, Z.-Y. Li and X.-H. Li, *J. Anal. At. Spectrom.*, 2021, **36**, 2063–2073.
46 J. Xie, D. Zhu, Q. Wang, L. Zhang, F. Cong, F. Nie, Y. Lu and L. Liu, *Geostandard Geoanalytic Res*, 2023, **47**, 143–154.
47 J. He, J. Meija and L. Yang, *Anal. Chem.*, 2021, acs.analchem.0c04744.
48 J. Meija, J. He, P. Grinberg, Z. Mester and L. Yang, *National Research Council of Canada*, , DOI:10.4224/CRM.2021.ZIRC-1.

49 J. Carignan, D. Cardinal, A. Eisenhauer, A. Galy, M. Rehkamper, F. Wombacher and N. Vigier, *Geostand Geoanalyt Res*, 2004, **28**, 139–148.
50 J. Vogl and W. Pritzkow, *J. Anal. At. Spectrom.*, 2010, **25**, 923.
51 F.-Z. Teng, N. Dauphas and J. M. Watkins, *Reviews in Mineralogy and Geochemistry*, 2017, **82**, 1–26.
52 W. A. Russell, D. A. Papanastassiou and T. A. Tombrello, *Geochimica et Cosmochimica Acta*, 1978, **42**, 1075–1090.
53 J. F. Minster and L. Ph. Ricard, *International Journal of Mass Spectrometry and Ion Physics*, 1981, **37**, 259–272.
54 G. J. Wasserburg, S. B. Jacobsen, D. J. DePaolo, M. T. McCulloch and T. Wen, *Geochimica et Cosmochimica Acta*, 1981, **45**, 2311–2323.
55 S. R. Hart and A. Zindler, *International Journal of Mass Spectrometry and Ion Processes*, 1989, **89**, 287–301.
56 M. F. Thirlwall, *Chemical Geology*, 1991, **94**, 85–104.
57 K. Habfast, *International Journal of Mass Spectrometry*, 1998, **176**, 133–148.
58 M. Rehkämper and A. N. Halliday, *International Journal of Mass Spectrometry*, 1998, **181**, 123–133.
59 B. Luais, P. Telouk and F. Albaréde, *Geochimica et Cosmochimica Acta*, 1997, **61**, 4847–4854.
60 N. S. Belshaw, P. A. Freedman, R. K. O'Nions, M. Frank and Y. Guo, *International Journal of Mass Spectrometry*, 1998, **181**, 51–58.
61 C. N. Marechal, P. Telouk and F. Albarede, *Chem Geol*, 1999, **156**, 251–273.
62 M. Rehkämper, M. Schönbächler and C. H. Stirling, *Geostandards and Geoanalytical Research*, 2001, **25**, 23–40.
63 F. Albarede, E. Albalat and P. Telouk, *Journal of Analytical Atomic Spectrometry*, , DOI:10.1039/C5JA00188A.
64 K. G. Heumann, S. M. Gallus, G. Rädlinger and J. Vogl, *J. Anal. At. Spectrom.*, 1998, **13**, 1001.
65 D. Vance and M. Thirlwall, *Chemical Geology*, 2002, **185**, 227–240.
66 F. Wombacher and M. Rehkämper, *J. Anal. At. Spectrom.*, 2003, **18**, 1371–1375.
67 F. Albarede, P. Telouk, J. Blichert-Toft, M. Boyet, A. Agranier and B. Nelson, *Geochim Cosmochim Ac*, 2004, **68**, 2725–2744.
68 M. F. Thirlwall and R. Anczkiewicz, *International Journal of Mass Spectrometry*, 2004, **235**, 59–81.
69 M. Nomura, K. Kogure and M. Okamoto, *International Journal of Mass Spectrometry and Ion Physics*, 1983, **50**, 219–227.
70 A. Quemet, C. Maillard and A. Ruas, *International Journal of Mass Spectrometry*, 2015, **392**, 34–40.
71 J. Zhang, N. Dauphas, A. M. Davis, I. Leya and A. Fedkin, *Nature Geosci*, 2012, **5**, 251–255.
72 R. C. J. Steele, T. Elliott, C. D. Coath and M. Regelous, *Geochimica et Cosmochimica Acta*, 2011, **75**, 7906–7925.
73 G. Budde, C. Burkhardt and T. Kleine, *Nat Astron*, 2019, **3**, 736–741.
74 G. Budde, G. J. Archer, F. L. H. Tissot, S. Tappe and T. Kleine, *J. Anal. At. Spectrom.*, 2022, **37**, 2005–2021.
75 H. L. Tang and N. Dauphas, *Earth Planet Sc Lett*, 2012, **359**, 248–263.

76 N. Dauphas and E. A. Schauble, *Annu Rev Earth Pl Sc*, 2016, **44**, 709–783.
77 N. X. Nie, N. Dauphas and R. C. Greenwood, *Earth and Planetary Science Letters*, 2017, **458**, 179–191.
78 Y. Di, Z. Li and Y. Amelin, *J. Anal. At. Spectrom.*, 2021, **36**, 1489–1502.
79 Y. Di, E. Krestianinov, S. Zink and Y. Amelin, *Chemical Geology*, 2021, **582**, 120411.
80 E. D. Young, A. Galy and H. Nagahara, *Geochimica et Cosmochimica Acta*, 2002, **66**, 1095–1104.

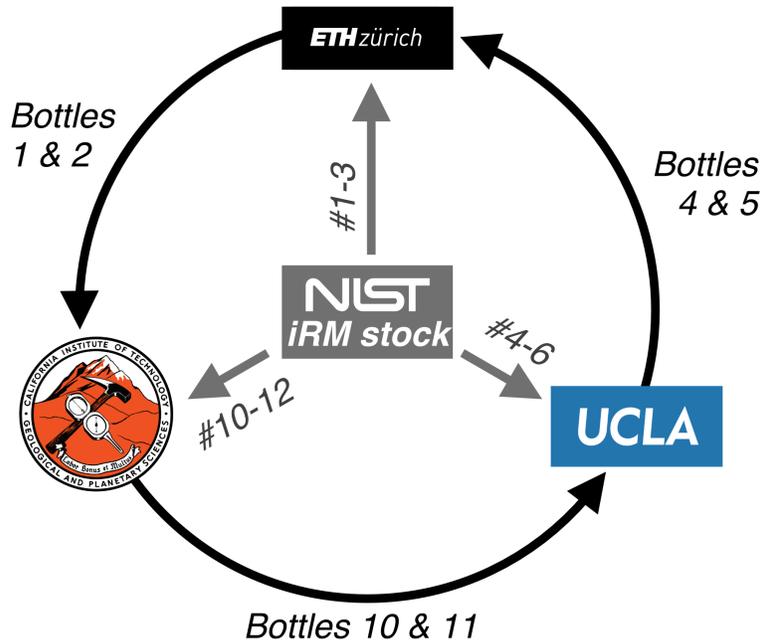

**Fig. 1.** Schematic depicting how the various iRM aliquots were distributed to and in between laboratories for the round-robin inter-calibration. To ensure data reproducibility and track sources of contamination, each lab analyzed 5 aliquots (3 shipped directly from NIST, and 2 from another lab after subsampling).

*1-column fitting image*

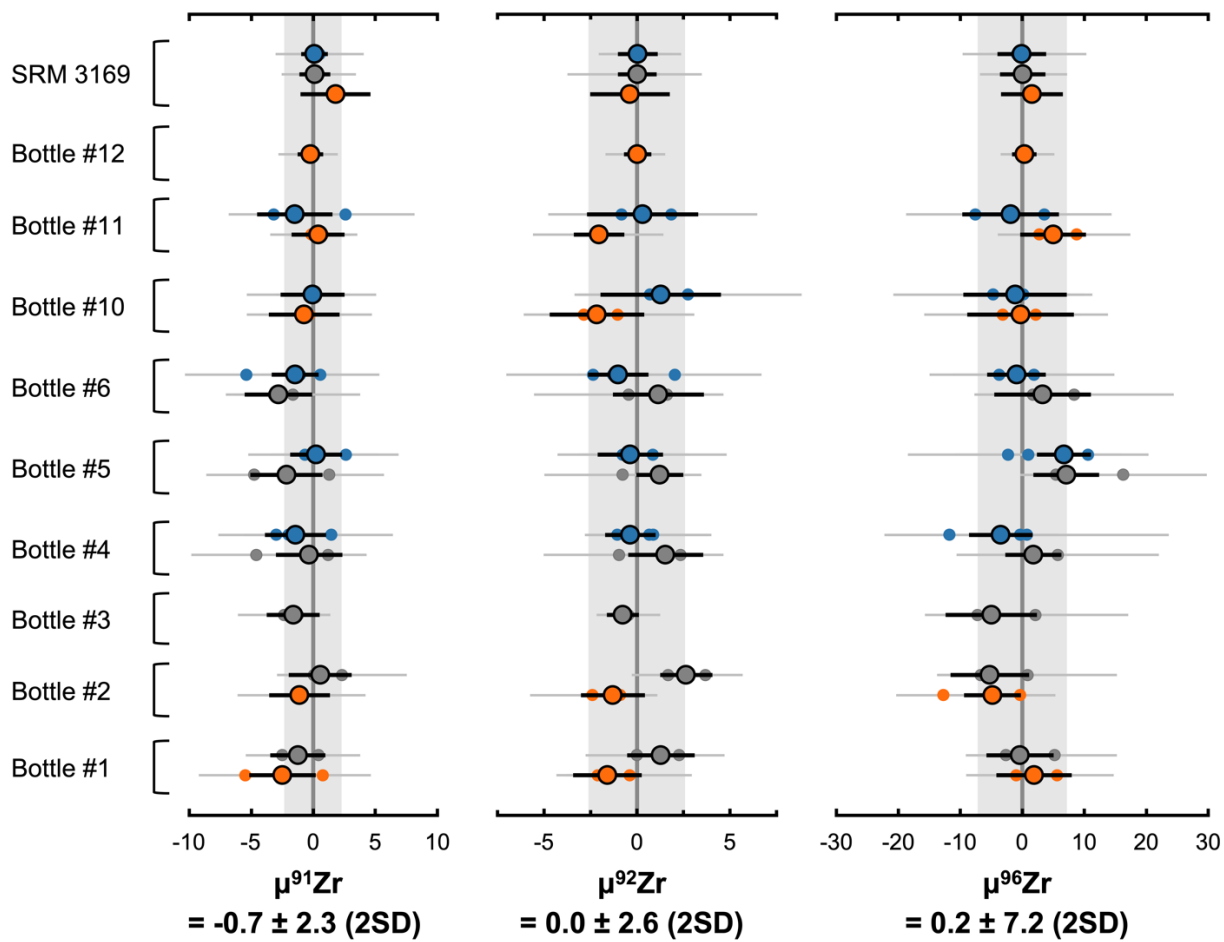

**Fig. 2.** Differences in Zr isotope composition (internally normalized to $^{94}Zr/^{90}Zr$) between the various aliquots of the candidate RM 8299 and SRM 3169, in 'µ' notation (ppm). Small symbols with grey errors bars are measurement session averages, while large symbols with black error bars show the averages of all measurement sessions from the same lab (Lab 1: orange; Lab 2: grey; Lab 3: blue). Uncertainties on each data point are 2 standard error (SE) as listed in Table 3. Data from Lab 1 are reported relative to bottle 12, whereas data from Lab 2 & 3 are reported relative to SRM 3169. The concordant data for all 3 laboratories confirm the isotopic homogeneity of the Zr candidate RM 8299, as well as the 2 lots of SRM 3169 measured, and the lack of contamination during the calibration process. Vertical gray bands represent the 2 standard deviation (SD) uncertainty of the average.

*2-column fitting image*

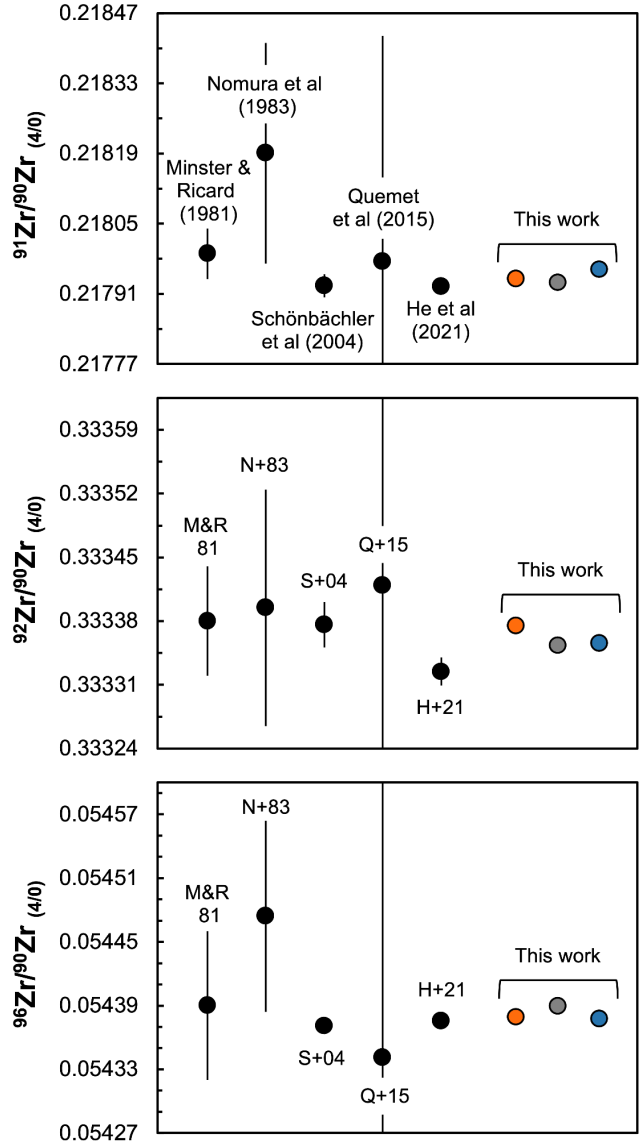

**Fig. 3.** Absolute Zr isotope ratios – all internally normalized using the exponential law to $^{94}Zr/^{90}Zr$ = 0.3381, following Ref.[53]. Colored symbols (Lab 1: orange; Lab 2: grey; Lab 3: blue) denote Zr isotope ratios for the candidate RM 8299 / SRM 3169 (this work), while black circles denote literature data obtained on unspecified Zr standards[53,69], multiple Zr standards including NIST SRM 3169 (Ref.[28]), a Zr single-element standard from SPEX[70], and ZIRC-1[47]. Despite good agreement with previous estimates, subtle differences are observed in the internally normalized data obtained in different laboratories on the very same iRM solution. (See *Section 6.2*). Uncertainties are 2SD. Source data in Table 4.

*1-column fitting image*

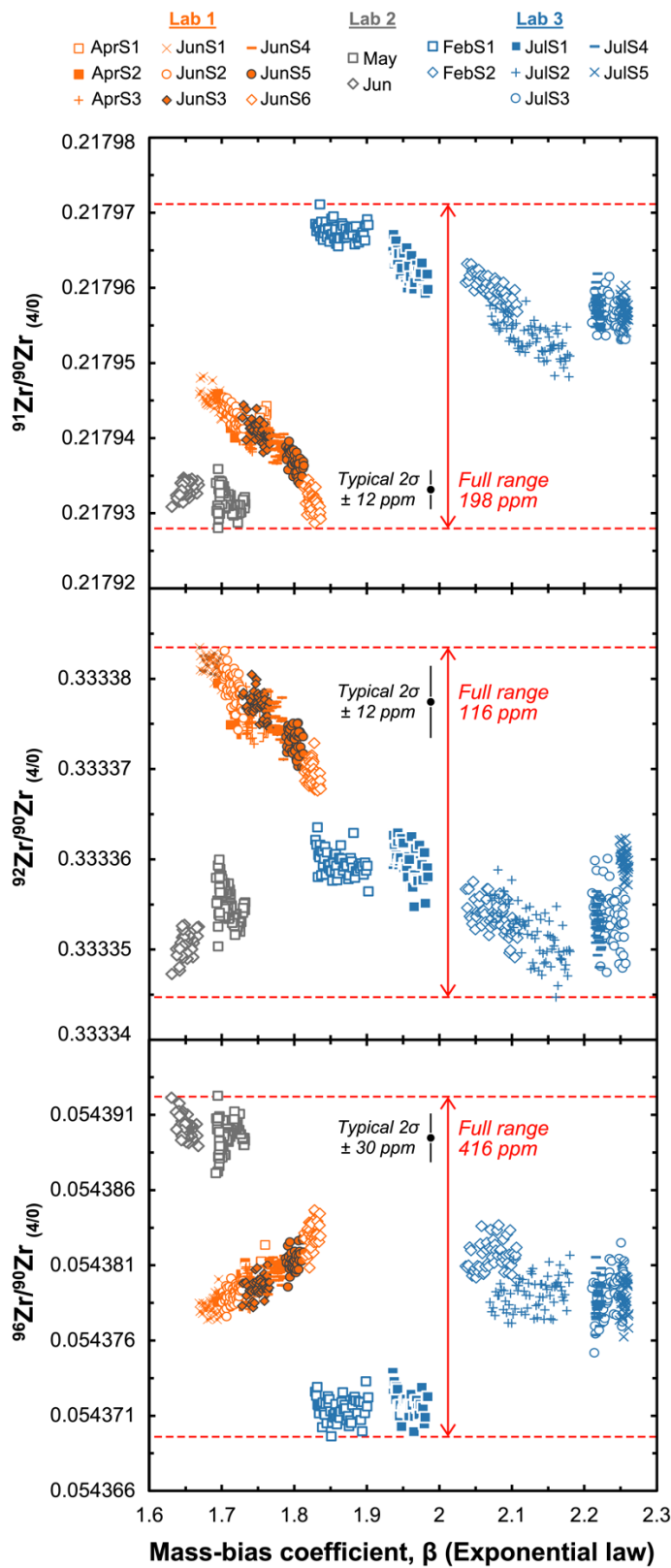

**Fig. 4.** Absolute Zr isotope ratios – internally normalized to $^{94}Zr/^{90}Zr = 0.3381$ using the exponential law (Eq. 2) for candidate RM 8299 and SRM 3169 – plotted as a function of the mass-bias coefficient, $β$. Despite mass-bias correction, differences in absolute ratios are observed (i) from session to session in any given laboratory, and (ii) between laboratories. Within a given laboratory, subtle to very pronounced trends are observed between internally normalized ratios and $β$ values. Both observations suggest these variations stem from a combination of non-exponential mass-dependent effects and mass-independent effects.

*1-column fitting image*

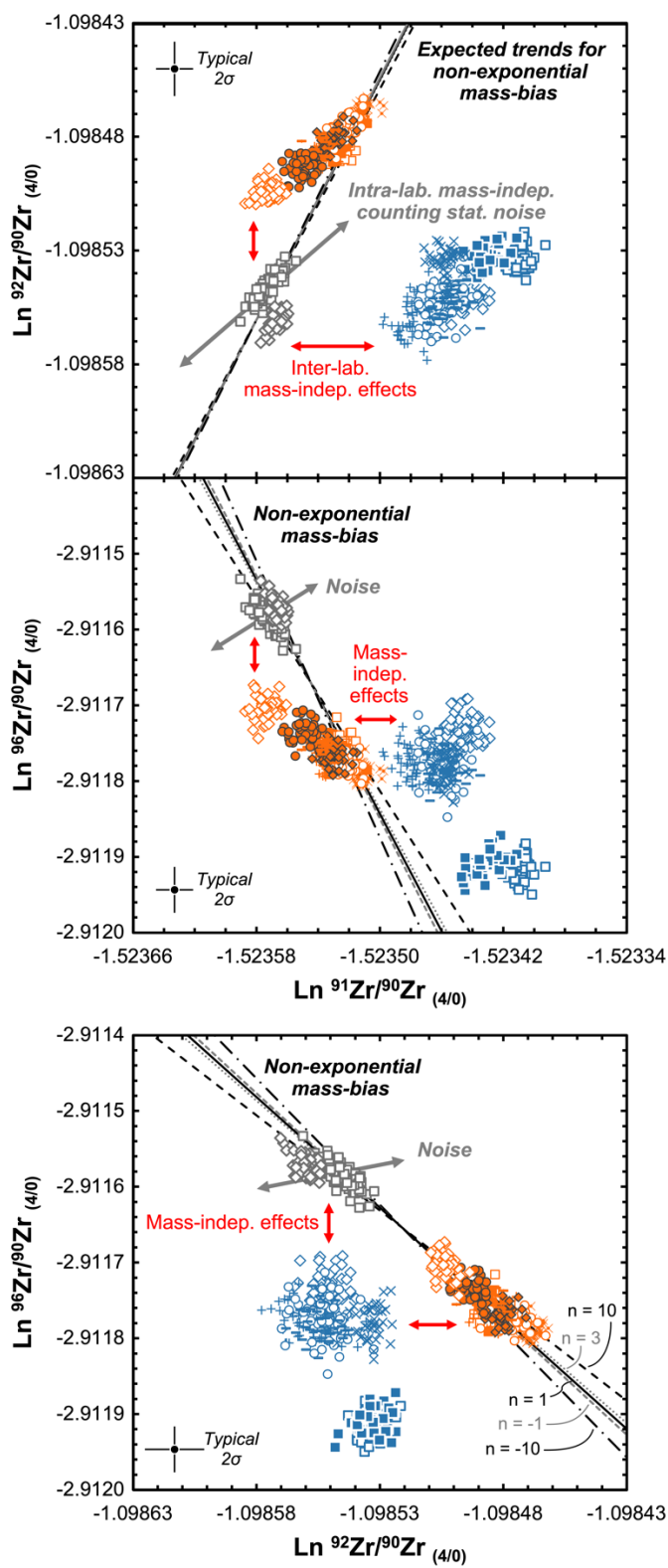

**Fig. 5.** Absolute Zr isotope ratios – internally normalized to $^{94}Zr/^{90}Zr$ = 0.3381 using the exponential law (Eq. 2) for candidate RM 8299 and SRM 3169 – shown in log-log triple isotope plots. Intra-laboratory trends appear to be controlled by a combination of non-exponential mass-dependent fractionation (dotted and dashed lines) and mass-independent counting statistics noise (grey arrows), primarily affecting $^{91}Zr$ (likely as a result of hydride formation: $^{90}ZrH^+$). Similarly, inter-laboratory offsets can be explained as a combination of non-exponential mass-dependent fractionation, and instrument-specific mass-independent effects. The latter is possibly related to cup degradation, or sample inlet system (red arrows). The slopes of the non-exponential mass-bias curves were calculated using Eq. (5) and GPL exponents n = -10, -1, +1, +3 and +10 (see label in bottom panels). The slopes of the noise lines *after internal normalization* (shown on the figure) were calculated by subtracting the slope of the exponential mass-fractionation line (Eq. 7) from the slope of the noise lines as described by Eq (63) in Ref.[67]). Symbols as on Figure 4.

*1-column fitting image*

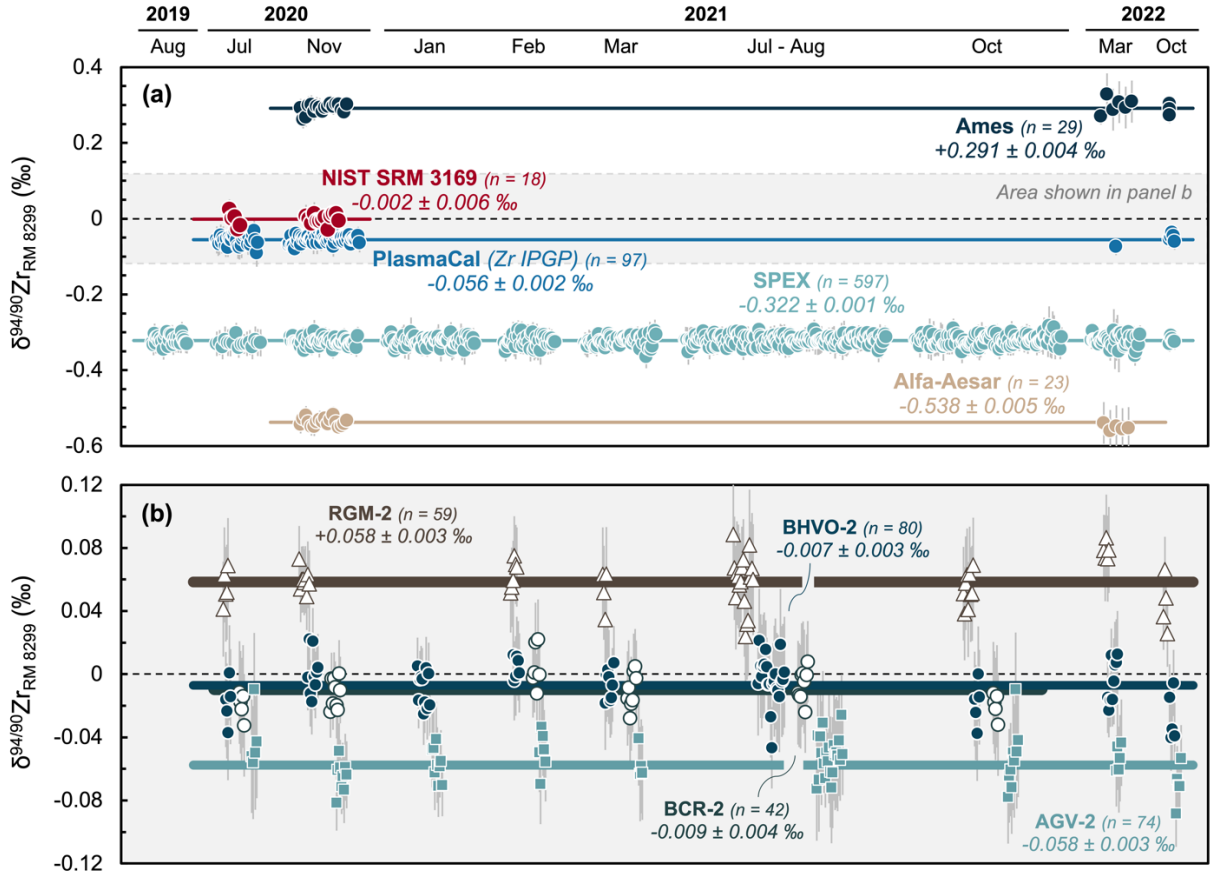

**Fig. 6.** $\delta^{94/90}$Zr values relative to RM 8299 for widely used Zr standards (a) and geostandards (b) used in the literature. Horizontal bars represent long-term average values, which were determined by repeat measurements carried out over more than 3 years (see dates on top of the figure) in Lab 1 using a $^{91}$Zr-$^{96}$Zr double spike. Uncertainties are 95 % CI. Data in Table 6.

*2-column fitting image*

**Table 1. Community identified guidelines for selection of an iRM.**

1) Must be demonstrably homogeneous.
2) Must be pure elements/chemical compounds easily dissolved into diluted acids.
3) Isotopic composition within the range of natural variability (ideally representative of a major geological reservoir).
4) Must be free of isotopic anomalies.
5) Community-led effort to choose the iRM against which to report data.
6) Must be widely available (i.e., conflict-of-interest free distribution).
7) Stock must be stable and sufficient to last decades.

*Recommendations from Teng et al. (2017)*

Table 2. Summary of the analytical setup and conditions used for Zr isotope analyses in each laboratory.

| Parameter | Lab 1 (Caltech) | Lab 2 (ETH) | Lab 3 (UCLA) |
|---|---|---|---|
| Mass spectrometer | Neptune Plus MC-ICPMS | Neptune Plus MC-ICPMS | Neptune MC-ICPMS |
| Sample introduction | Wet plasma (glass spray chamber) | Dry plasma (Aridus II) | Wet plasma (glass spray chamber) |
| Nebulizer type / flow | PFA nebulizer, 50 µl/min | PFA nebulizer, 50 µl/min | PFA nebulizer, 50 µl/min |
| Cones | General H sampler and H skimmer | General H sampler and H skimmer | General H sampler and H skimmer |
| Resolution | Low resolution | Low resolution | Low resolution |
| Cup Configuration | L4 / L3 / L2 / L1 / **C** / H1 / H2 / H3 / H4 | L4 / L3 / L2 / L1 / **C** / H1 / H2 / H3 / H4 | L3 / L2 / L1 / **C** / H1 / H2 / H4 |
| Isotopes monitored | $^{90}$Zr, $^{91}$Zr, $^{92}$Zr, $^{93}$Nb, **$^{94}$Zr**, $^{95}$Mo, $^{96}$Zr, $^{97}$Mo, $^{98}$Mo | $^{90}$Zr, $^{91}$Zr, $^{92}$Zr, $^{94}$Zr, **$^{95}$Mo**, $^{96}$Zr, $^{99}$Ru, $^{100}$Mo/Ru, $^{101}$Ru | $^{177}$Hf$^{++}$, $^{49}$Ti$^{40}$Ar, $^{90}$Zr, **$^{91}$Zr**, $^{92}$Zr, $^{94}$Zr, $^{96}$Zr |
| Amplifiers | 10$^{12}$ Ω on H1 & H4; 10$^{11}$ Ω on all other cups | 10$^{12}$ Ω on H2 & H4; 10$^{11}$ Ω on all other cups | 10$^{11}$ Ω on all cups |
| Gain calibration | Daily | Daily | Weekly |
| Acid matrix | 0.58 M HNO$_3$ + 0.01 M HF | 0.5 M HNO$_3$ + 0.005 M HF | 0.3 M HNO$_3$ + 0.005 M HF |
| Solution [Zr] | 1 µg/g | 350 ng/g | 2 µg/g |
| Intensity on $^{90}$Zr | 22-23 V | 42-47 V | 11-29 V |
| Instrument sensitivity | 43-45 V / ppm of Zr | 233-261 V / ppm of Zr | 11-28 V / ppm of Zr |
| Uptake time | 55 s | 55 s | 100 s |
| Measurement | 1 block of 50 cycles | 3 blocks of 20 cycles | 1 block of 20 cycles |
| Cycle duration | 4.194 s | 4.194 s | 8.388 s |
| On peak zero | 10 cycles | 3 blocks of 20 cycles | 1 block of 20 cycles |
| Cycle duration | 4.194 s | 4.194 s | 8.388 s |
| Rinse duration | 568 s | 310 s | 150 s |

Bold font in the "Cup configuration" and "Isotopes monitored" rows denotes the central cup.

Table 3. Differences in Zr isotope composition of the various RM 8299 aliquots and SRM 3169 lots.

| Sample | Lot # | Session | Seq. | n | µ91Zr | 2SE | µ92Zr | 2SE | µ96Zr | 2SE | Rel. to |
|---|---|---|---|---|---|---|---|---|---|---|---|
| *Laboratory 1* | | | | | | | | | | | |
| Bottle 1 | | Apr | 1 | 9 | -5.5 | 3.7 | -0.4 | 3.3 | 5.6 | 9.2 | Bottle 12 |
| Bottle 1 | | June | 1 | 10 | 0.8 | 3.9 | -2.1 | 2.2 | -1.0 | 8.1 | Bottle 12 |
| ***Bottle 1 avg*** | | | | | ***-2.5*** | ***2.7*** | ***-1.6*** | ***1.8*** | ***1.9*** | ***6.1*** | |
| Bottle 2 | | Apr | 1 | 8 | -0.9 | 5.2 | -2.4 | 3.4 | -12.7 | 7.6 | Bottle 12 |
| Bottle 2 | | June | 1 | 9 | -1.1 | 2.8 | -0.9 | 2.0 | -0.3 | 5.7 | Bottle 12 |
| ***Bottle 2 avg*** | | | | | ***-1.1*** | ***2.5*** | ***-1.3*** | ***1.7*** | ***-4.8*** | ***4.6*** | |
| Bottle 10 | | Apr | 1 | 8 | -0.3 | 5.0 | -2.9 | 3.2 | -3.1 | 12.7 | Bottle 12 |
| Bottle 10 | | June | 1 | 9 | -0.9 | 3.5 | -1.0 | 4.1 | 2.2 | 11.7 | Bottle 12 |
| ***Bottle 10 avg*** | | | | | ***-0.7*** | ***2.9*** | ***-2.2*** | ***2.5*** | ***-0.3*** | ***8.6*** | |
| Bottle 11 | | Apr | 1 | 8 | -0.1 | 3.3 | -2.0 | 1.5 | 8.8 | 8.7 | Bottle 12 |
| Bottle 11 | | June | 1 | 9 | 0.8 | 2.8 | -2.1 | 3.5 | 2.8 | 6.7 | Bottle 12 |
| ***Bottle 11 avg*** | | | | | ***0.4*** | ***2.1*** | ***-2.0*** | ***1.4*** | ***5.0*** | ***5.3*** | |
| Bottle 12 | | Apr | 1 | 34 | -0.4 | 2.4 | -0.1 | 1.6 | 0.9 | 4.3 | Bottle 12 |
| Bottle 12 | | June | 1 | 151 | -0.2 | 1.1 | 0.1 | 0.8 | 0.2 | 2.2 | Bottle 12 |
| ***Bottle 12 avg*** | | | | | ***-0.2*** | ***1.0*** | ***0.0*** | ***0.7*** | ***0.3*** | ***2.0*** | |
| SRM 3169 | 130920 | June | 3 | 21 | 1.8 | 2.8 | -0.4 | 2.1 | 1.6 | 5.0 | Bottle 12 |
| | | | | | | | | | | | |
| *Laboratory 2* | | | | | | | | | | | |
| Bottle 1 | | May | 1 | 6 | -2.5 | 2.9 | 0.0 | 2.8 | 5.2 | 10.1 | SRM 3169 |
| Bottle 1 | | June | 2 | 6 | 0.4 | 3.4 | 2.3 | 2.4 | -2.6 | 6.5 | SRM 3169 |
| ***Bottle 1 avg*** | | | | | ***-1.2*** | ***2.2*** | ***1.3*** | ***1.8*** | ***-0.4*** | ***5.4*** | |
| Bottle 2 | | May | 1 | 6 | 2.3 | 5.2 | 1.7 | 2.0 | -6.7 | 7.0 | SRM 3169 |
| Bottle 2 | | June | 2 | 6 | 0.0 | 2.9 | 3.7 | 2.0 | 0.9 | 14.4 | SRM 3169 |
| ***Bottle 2 avg*** | | | | | ***0.6*** | ***2.5*** | ***2.6*** | ***1.4*** | ***-5.2*** | ***6.3*** | |
| Bottle 3 | | May | 1 | 6 | -1.2 | 2.6 | -0.9 | 1.0 | -7.2 | 8.4 | SRM 3169 |
| Bottle 3 | | June | 2 | 6 | -2.3 | 3.7 | -0.5 | 1.7 | 2.1 | 15.0 | SRM 3169 |
| ***Bottle 3 avg*** | | | | | ***-1.6*** | ***2.1*** | ***-0.8*** | ***0.9*** | ***-5.0*** | ***7.4*** | |
| Bottle 4 | | May | 1 | 6 | 1.2 | 3.1 | 2.3 | 2.3 | 1.5 | 4.7 | SRM 3169 |
| Bottle 4 | | June | 2 | 6 | -4.6 | 5.2 | -1.0 | 4.1 | 5.7 | 16.3 | SRM 3169 |
| ***Bottle 4 avg*** | | | | | ***-0.3*** | ***2.7*** | ***1.5*** | ***2.0*** | ***1.8*** | ***4.5*** | |
| Bottle 5 | | May | 1 | 6 | 1.3 | 4.4 | 1.4 | 1.3 | 5.4 | 5.8 | SRM 3169 |
| Bottle 5 | | June | 2 | 6 | -4.8 | 3.9 | -0.8 | 4.2 | 16.3 | 13.5 | SRM 3169 |
| ***Bottle 5 avg*** | | | | | ***-2.1*** | ***2.9*** | ***1.2*** | ***1.3*** | ***7.1*** | ***5.3*** | |
| Bottle 6 | | May | 1 | 6 | -1.6 | 5.4 | -0.5 | 5.1 | 8.4 | 16.1 | SRM 3169 |
| Bottle 6 | | June | 2 | 6 | -3.2 | 3.1 | 1.6 | 2.8 | 1.7 | 8.9 | SRM 3169 |
| ***Bottle 6 avg*** | | | | | ***-2.8*** | ***2.7*** | ***1.1*** | ***2.4*** | ***3.3*** | ***7.8*** | |
| SRM 3169 | 071226 | May | 1 | 37 | 0.1 | 1.4 | 0.0 | 1.1 | 0.0 | 4.3 | SRM 3169 |
| SRM 3169 | 071226 | June | 2 | 7 | 0.5 | 3.0 | -0.1 | 3.6 | 0.2 | 7.0 | SRM 3169 |
| ***SRM 3169 avg*** | | | | | ***0.1*** | ***1.3*** | ***0.0*** | ***1.0*** | ***0.1*** | ***3.7*** | |
| | | | | | | | | | | | |
| *Laboratory 3* | | | | | | | | | | | |
| Bottle 4 | | Feb | 1 | 18 | -2.0 | 3.6 | -1.1 | 1.7 | -0.4 | 6.3 | SRM 3169 |
| Bottle 4 | | Jul 1 | 1 | 18 | 1.5 | 5.0 | 0.7 | 3.3 | -11.7 | 9.8 | SRM 3169 |
| Bottle 4 | | Jul 2 | 5 | 5 | -3.0 | 4.7 | 0.9 | 2.9 | 0.7 | 22.9 | SRM 3169 |
| ***Bottle 4 avg*** | | | | | ***-1.4*** | ***2.5*** | ***-0.4*** | ***1.4*** | ***-3.5*** | ***5.1*** | |
| Bottle 5 | | Feb | 1 | 18 | -0.6 | 2.7 | -0.6 | 2.3 | -2.3 | 8.5 | SRM 3169 |
| Bottle 5 | | Jul 1 | 1 | 16 | 0.0 | 5.2 | 0.8 | 4.0 | 10.6 | 5.3 | SRM 3169 |
| Bottle 5 | | Jul 2 | 5 | 5 | 2.6 | 4.3 | -0.8 | 3.5 | 1.0 | 19.4 | SRM 3169 |
| ***Bottle 5 avg*** | | | | | ***0.2*** | ***2.1*** | ***-0.4*** | ***1.8*** | ***6.7*** | ***4.4*** | |
| Bottle 6 | | Feb | 1 | 18 | -1.1 | 2.3 | -1.3 | 1.9 | -0.7 | 5.7 | SRM 3169 |
| Bottle 6 | | Jul 1 | 2 | 15 | -5.4 | 4.9 | 2.0 | 4.6 | 1.9 | 13.0 | SRM 3169 |
| Bottle 6 | | Jul 2 | 4 | 9 | 0.6 | 4.8 | -2.4 | 4.7 | -3.8 | 11.2 | SRM 3169 |
| ***Bottle 6 avg*** | | | | | ***-1.5*** | ***1.9*** | ***-1.0*** | ***1.6*** | ***-0.9*** | ***4.7*** | |
| Bottle 10 | | Jul 1 | 2 | 14 | 0.0 | 3.0 | 2.7 | 6.1 | 0.2 | 9.7 | SRM 3169 |
| Bottle 10 | | Jul 2 | 5 | 9 | -0.1 | 5.2 | 0.7 | 3.8 | -4.7 | 16.0 | SRM 3169 |
| ***Bottle 10 avg*** | | | | | ***0.0*** | ***2.6*** | ***1.3*** | ***3.2*** | ***-1.1*** | ***8.3*** | |
| Bottle 11 | | Jul 1 | 2 | 16 | -3.2 | 3.6 | 1.8 | 4.6 | -7.6 | 11.2 | SRM 3169 |
| Bottle 11 | | Jul 2 | 4 | 9 | 2.6 | 5.6 | -0.8 | 3.9 | 3.5 | 10.9 | SRM 3169 |
| ***Bottle 11 avg*** | | | | | ***-1.5*** | ***3.0*** | ***0.3*** | ***3.0*** | ***-1.9*** | ***7.8*** | |
| SRM 3169 | 130920 | Feb | 1 | 52 | 0.0 | 1.3 | 0.1 | 1.7 | 0.5 | 6.1 | SRM 3169 |
| SRM 3169 | 130920 | Jul 1 | 1 | 77 | 0.2 | 2.5 | -0.1 | 1.7 | -0.9 | 6.1 | SRM 3169 |
| SRM 3169 | 130920 | Jul 2 | 4 | 35 | 0.5 | 3.6 | 0.1 | 2.2 | 0.4 | 10.0 | SRM 3169 |
| ***SRM 3169 avg*** | | | | | ***0.1*** | ***1.1*** | ***0.0*** | ***1.1*** | ***-0.1*** | ***3.9*** | |

**Table 4. Zirconium isotope composition of NIST RM 8299, compared to literature values for natural Zr.**

|  | $^{91}Zr/^{90}Zr$ | 2SD | $^{92}Zr/^{90}Zr$ | 2SD | $^{94}Zr/^{90}Zr$ | $^{96}Zr/^{90}Zr$ | 2SD |
|---|---|---|---|---|---|---|---|
| Natural Zr - M&R81 | 0.21799 | 0.00005 | 0.33338 | 0.00006 | 0.3381 | 0.05439 | 0.00007 |
| Natural Zr - N+83 | 0.21819 | 0.00022 | 0.33339 | 0.00013 | 0.3381 | 0.05447 | 0.00009 |
| Several Zr standards - S+04 | 0.217926 | 0.000023 | 0.333376 | 0.000025 | 0.3381 | 0.054371 | 0.000008 |
| SPEX Zr standard - Q+15 | 0.21797 | 0.00045 | 0.33342 | 0.00079 | 0.3381 | 0.05434 | 0.00050 |
| ZIRC-1* | 0.217925 | 0.000014 | 0.333324 | 0.000016 | 0.3381 | 0.054375 | 0.000006 |
| NIST RM 8299 - Lab 1 | 0.217940 | 0.000008 | 0.333376 | 0.000007 | 0.3381 | 0.054380 | 0.000003 |
| NIST RM 8299 - Lab 2 | 0.217932 | 0.000003 | 0.333354 | 0.000006 | 0.3381 | 0.054390 | 0.000002 |
| NIST RM 8299 - Lab 3 | 0.217959 | 0.000010 | 0.333355 | 0.000008 | 0.3381 | 0.054377 | 0.000008 |
| *NIST RM 8299 - Avg* | *0.217944* | *0.000027* | *0.333362* | *0.000025* | *0.3381* | *0.054382* | *0.000013* |

*Values calculated from the raw data (corrected for OPZ and Mo interferences) in the supplementary of He et al (2021).

All data are internally normalized to $^{94}Zr/^{90}Zr$ = 0.3381 (after Minster & Ricard 1981).

M&R81 = Minster & Ricard (1981); N+83 = Nomura et al. (1983); S+04 = Schönbächler et al. (2004); Q+15 = Quemet et al. (2015).

**Table 5. How available Zr reference materials fulfill the community-identified guidelines for an iRM.**

| Reference material | SRM 3169 / RM 8299 | IPGP-Zr | ZIRC-1 | Zr solution |
|---|---|---|---|---|
| Manufacturer | NIST | PlasmaCal | NRC | Alfa-Aesar |
| 1) Demonstrated homogeneity | ✓ | ≈* | ≈* | ≈* |
| 2) Pure element easily dissolved into diluted acids | ✓ | ✓ | ✓ | ✓ |
| 3) $\delta^{94/90}$Zr representative of a major geological reservoir | ✓ | | | |
| 4) Must be free of isotopic anomalies | ✓ | TBD | TBD | |
| 5a) Choice is a community-led effort | ✓ | | | |
| 5b) Wide acceptance (number of publications using this standard) | 26** | 16 | 3 | 5 |
| 6) Conflict-of-interest free distribution | ✓ | | ✓ | |
| 7) Stock must be stable and sufficient to last decades | ✓ | | ✓ | |

*While homogeneity should be ensured by the fact that this proposed RM exist in solution form, to the best of our knowledge, no bottle-to-bottle isotopic homogeneity study has been performed.

** Including 16 publications focused on mass-dependent isotope effects (Refs. 7-9,11,12,26,29-38), and 10 focused on mass-independent effects (Refs. 13,15-23).

Table 6. $\delta^{94/90}$Zr values (‰, relative to RM 8299) of widely used Zr standards and geostandards.

| Material | Information | Lot # | $\delta^{94/90}$Zr$_{RM\ 8299}$ | 2SD | 2SE | MSWD | n | Ref. |
|---|---|---|---|---|---|---|---|---|
| **_Pure Zr standard_** | | | | | | | | |
| Alfa-Aesar | | 03-14247H | -0.538 | 0.022 | 0.005 | 0.734 | 23 | This work |
| Ames | | | 0.291 | 0.028 | 0.004 | 1.163 | 29 | This work |
| IPGP Zr | _Plasma_Cal SCP Science | 5131203028 | -0.056 | 0.023 | 0.002 | 0.939 | 98 | This work |
| SPEX | CertiPrep, 'Assurance' grade | 21-168ZRM | -0.322 | 0.025 | 0.001 | 0.163 | 597 | This work |
| SRM 3169 | NIST Zr Standard Solution | 130920 | -0.002 | 0.029 | 0.006 | 1.257 | 18 | This work |
| ZIRC-1* | Zr CRM from the NRC | | -0.276 | 0.033 | 0.006 | | 67 | [1] |
| **_Zircon standards_** | | | | | | | | |
| GJ-1** | From African pegmatites | | -0.064 | 0.032 | | | 12 | [2] |
| MTUR1 | Mud Tank carbonatite | | -0.055 | 0.028 | 0.002 | 1.010 | 151 | [3] |
| 91500 | Kuel Lake, Ontario, Canada | | -0.135 | 0.037 | 0.013 | 0.959 | 8 | [3] |
| **_Other geostandards_** | | | | | | | | |
| AGV-2 | Andesite | | -0.058 | 0.028 | 0.003 | 1.122 | 74 | This work |
| BCR-2 | Basalt, Columbia River | | -0.009 | 0.024 | 0.004 | 0.922 | 42 | This work |
| BHVO-2 | Basalt, Hawaiian Volcanic Observatory | | -0.007 | 0.029 | 0.003 | 1.579 | 80 | This work |
| RGM-2 | Rhyolite, Glass Mountain | | 0.058 | 0.028 | 0.003 | 1.046 | 59 | This work |

References: [1] Tian et al. (2022); [2] Tian et al. (2020); [3] Tompkins et al. (2020).

*$\delta^{94/90}$Zr$_{RM\ 8299}$ calculated from the reported $\delta^{94/90}$Zr of -0.220 ± 0.023/0.006 ‰ (2SD/2SE) of ZIRC-1 relative to IPGP-Zr, and the IPGP-Zr composition relative to RM 8299. Errors were added in quadrature. Note that _n_ is reported as 7 in the original publication (Tian et al. 2022), but the true number of measurements used in the study is 67 (see their Table 2).

*$\delta^{94/90}$Zr$_{RM\ 8299}$ calculated from the reported $\delta^{94/90}$Zr of -0.008 ± 0.022 ‰ (2SD) of GJ-1 relative to IPGP-Zr, and the IPGP-Zr composition relative to RM 8299. Errors were added in quadrature. Note that _n_ is reported as 3 in the original publication (Tian et al. 2020), but the true number of measurements used in the study is 12 (see their Table 1).